\documentclass{article}

\usepackage{microtype}
\usepackage{graphicx}
\usepackage{booktabs}
\usepackage{hyperref}

\ifdefined\ANON
\usepackage{icml2026}
\else
\usepackage[preprint]{icml2026}
\fi

\usepackage{amsmath}
\usepackage{amssymb}
\usepackage[capitalize,noabbrev]{cleveref}

\icmltitlerunning{What moves the ceiling on automated audit, from code to science}

\begin{document}

\twocolumn[
\icmltitle{What Moves the Ceiling on Automated Audit:\\
           from General Code to Scientific Code, Results and Data}

\begin{icmlauthorlist}
\icmlauthor{Zhaohe Dong}{ind}
\end{icmlauthorlist}

\icmlaffiliation{ind}{Independent researcher}
\icmlcorrespondingauthor{Zhaohe Dong}{dongzhaohe321418@gmail.com}

\icmlkeywords{code review, LLM critics, AI for science, evaluation, preregistration}

\vskip 0.3in
]

\printAffiliationsAndNotice{}

\begin{abstract}
Models increasingly check what other models write. We study one shipped arrangement, in which an
auditor from a different vendor reads the work and deterministic checks verify what code can verify,
and measure its limits against ground truth no model supplies. On general-purpose code with hidden
tests, eight readings of the auditor reach 30.0\% recall [20.0, 40.7] at 16.0\% [10.1, 22.3] flags on
correct code, with diminishing returns and no plateau. The reading model (confounded with its sampling), one
rulebook sentence and the decision rule each move that figure, and much of what is missed is, on a
post hoc rating, behaviour the specification never determined. In three preregistered science
studies, checking that reported numbers trace to their sources flagged all 49 inconsistent reports
and none of 25 consistent fabrications, of which the model auditor flagged 23 (21 with a sound
finding, post hoc); on data, a documentation-derived validator and the auditor flagged overlapping but different
faulty items, together 112 of 140 at a 2.9\% [0.7, 5.7] flag rate on clean data; and on scientific code,
reframing the task around one step cut flags on correct code from 68.7\% to 36.0\%. Where both
tiers ran, the deterministic one flagged no clean item and could not see consistent error; the model
saw some and erred itself. Studies making model calls were preregistered, the two unregistered
analyses are named, and every quoted report passed independent cross-vendor review.
\end{abstract}

\section{Introduction}
When a model writes code, reports a result or prepares a dataset, someone has to check it, and
increasingly that someone is another model. CrossAudit, the system every auditor reading here passes
through (Section~\ref{sec:system}), is built on two rules for doing that. The model that audits must come
from a different vendor from the model that wrote the work, and what a deterministic check can verify
is verified by code, not by a model's opinion. An auditor of this kind reads work it did not write
and says whether to block it. We ask how much it catches, what decides the limit, and whether the
arrangement helps where checking matters most, in science.

A reviewer's score on a benchmark mixes at least four things: whether the model can see a defect,
whether it judges the defect worth blocking on, whether the benchmark's hidden tests define the
behaviour it would have to see, and how many times it is allowed to look. Work on critic models
improves the first two and reports the aggregate gain. We vary these factors one at a time where the
design allows it, against a truth no model supplies: a hidden test suite executed by an interpreter.
Some contrasts cannot be isolated: changing the reading model also changed its sampling, and the
role of the specification is a post hoc association.

\textbf{The limit on general-purpose code} (Section~\ref{sec:ceiling}). Eight independent readings of the
shipped cross-vendor auditor reach 30.0\% union recall [20.0, 40.7] on defective code at 16.0\%
[10.1, 22.3] on correct code, and the eighth reading still adds 1.93 points [1.14, 2.78], so returns diminish
without a plateau. A stronger same-vendor model, sampled differently, moves recall by $-26.4$
points [$-37.3$, $-15.6$] while flagging 3.3\% [0.7, 6.7] of correct code against 16.0\%, and the contrast
reverses sign under a weaker, exploratory any-finding rule. One added rulebook sentence moves flags at one reading on 56 defective instances by $+26.8$
points [13.6, 40.4] (an exploratory flag contrast), at $+12.5$ points on correct code (an interval,
[$-3.6$, $+26.8$], that does not exclude zero), without measurably moving outcomes. Of the 57
defective instances no reading of the first three auditor families flagged, two model raters judge 44
(77.2\% [62.1, 91.1], post hoc) to be behaviour the specification never determined. A recall figure is therefore contingent on auditor route, rulebook, decision rule and benchmark,
and is not portable without its false-positive rate.

\textbf{Auditing science} (Section~\ref{sec:ai4s}). We then take the system into science, with one
preregistered study for each job a scientist would hand an auditor. On reported results, the
deterministic check that every number traces to its source flagged all 49 inconsistencies among a
report, its results file and its run log, and none of 25 results that were wrong consistently in all
three; the model auditor flagged 23 of those 25, 21 with a sound finding about the fault (post hoc).
On data, a documentation-derived validator and the auditor flagged overlapping but different
faulty items, 92 and 65 of 140, and together 112, at the auditor's 2.9\% flag rate on clean data
(post hoc, counting only findings about the injected fault: 61 and 110). On scientific code with no
tests shown, reframing the task around the step rather than the whole problem cut the auditor's flags
on correct code from 68.7\% to 36.0\% and its recall from 87.7\% to 72.8\%, on the same programs.

\textbf{Contributions.}
\begin{itemize}\itemsep1pt
\item A measurement programme for an automated auditor's limit against executed ground truth, with
  union recall always reported beside the false-positive rate it cost: controlled changes to the
  number of readings and the rulebook, route comparisons that change model and sampling together,
  re-grading under decision rules, and a post hoc association with what the specification determines.
\item Evidence that no single factor fixes the limit on general-purpose code: the number of
  readings, the rulebook, the route and the decision rule each move it, and what is missed is
  associated (post hoc, not isolated as a cause) with what the specification leaves undetermined; a
  recall figure without its operating point does not transfer.
\item Operating points of the model auditor on scientific code, and comparisons of the model tier with
  deterministic verification on reported results (the shipped provenance checks and re-execution) and
  data (a documentation-derived validator), against executed or constructed ground truth. The tiers
  flag different faults, and each fails where the other does not.
\item A methodological finding: in the results and data studies, reading every flag on a clean item
  showed the auditor flagging real defects in our own construction of those items, so an evaluation of
  an auditor must establish that its clean items are clean by the auditor's standard, or its
  false-positive rate can measure the evaluator.
\end{itemize}

\textbf{Preregistration and review.} Every study that made model calls was preregistered before its
first one, with hypotheses, intervals and stopping rules fixed in advance, and planned departures
recorded as numbered amendments. Two whole analyses were not registered in advance, and the paper names
both (Appendix~\ref{sec:method}); every other unregistered quantity is labelled post hoc where it
appears. Four preregistered hypotheses were killed by their own rules, and
in the science studies one more. Every report was reviewed by a model from a different vendor with
access to the raw records, at between three and twenty-one rounds each (the derived-quantities report within the
whole-manuscript review), and the reviews are archived verbatim. Section~\ref{sec:discussion} discusses what the measurements mean. The full method, results
and discussion are in Appendices~\ref{sec:method}--\ref{sec:discussion-full}, and the records, reviews
and the command that regenerates every figure are described in Appendix~\ref{sec:repro}.

\section{The system under study}
\label{sec:system}

Every auditor reading in this paper is taken through one shipped system, CrossAudit. Every auditor
prompt is built by the product's own prompt builder, and every family but one is read through
the product's own audit call; the exception, \texttt{gpt-6-astra}, receives the same prompt
bytes through a direct call. We describe the system first because
its design fixes what the later sections can and cannot vary. We state its structure only here
and quote no measurement of it. Its limits are the subject of
Section~\ref{sec:ceiling} onward.

\textbf{Separation of roles.} A generator model writes an increment (a set of files) into a Git
repository. An auditor model then reads the committed bytes and returns findings, each graded
BLOCKER or ADVISORY. The auditor must come from a different vendor from the generator: the audit
refuses to run when the configuration names the same vendor for both roles, and when the
generator's vendor is not declared, the receipt records that independence could not be asserted.
The same-vendor arms below run with the generator's vendor undeclared, deliberately. The premise is that a model
reviewing its own vendor's output is an interested reader. The product takes that premise as given, and Section~\ref{sec:ceiling} puts
one consequence of it to the test: whether a same-vendor auditor catches more or less.

\textbf{Two tiers of evidence.} Before the auditor reads anything, a deterministic check layer
runs registered checks over the same committed bytes. A check's finding is \emph{verified}: code
produced it and code can reproduce it. A model's finding is \emph{raised}: it is a reading, not
yet reproduced. The checks are grouped into named profiles, so strictness is a setting the
project chooses:
\begin{itemize}\itemsep0pt
\item \texttt{general}, the default: structured files parse, every local path a file lists as an
  input or dependency exists, and relative links and leftover placeholders are reported as
  advisory.
\item \texttt{science}: results carry a schema; quantities carry units; a result that reports
  convergence must meet its own threshold; every input a result depends on is listed, exists and is cited at an exact revision
  (\emph{provenance}); and every number a report states is annotated with the file and quoted span
  it came from, which a check matches against the cited bytes (\emph{number$\to$source}).
\item \texttt{research}: the general profile's checks plus a citation check under which a report may cite only
  sources it fetched through a governed tool.
\end{itemize}
The science profile follows one rule throughout, and the rule is the one Section~\ref{sec:ai4s} tests: \emph{the
model names its evidence, and code verifies it.} A number with no source line, a unit that
disagrees with its schema or an input that does not exist fails without any model's opinion.

\textbf{Computed verdicts.} The verdict is derived by code from the evidence, in a
fixed order. A failed deterministic check blocks, whatever the model says. A model BLOCKER with
no deterministic support either blocks or escalates to a person, according to a project setting.
A blocked increment goes back to the generator for a bounded number of revision rounds, three by
default. Each revision passes a repair guard before it is committed: an edit outside the audited
scope is refused, and a new catch-all handler, a deleted assertion or a skipped test is flagged to
the auditor for the next round.

\textbf{Receipts.} A passing round produces a receipt that binds the verdict to the commit that was
audited, the commit of the rulebook the auditor was given, and a digest over every finding with its
tier. Every model call is written to a ledger with its prompt hash and cost. A verification command
recomputes the digest from the recorded evidence, so a receipt can be checked by someone who does
not trust the operator.

\textbf{What the rest of the paper measures.} Section~\ref{sec:ceiling}, with
Appendices~\ref{sec:method} and~\ref{sec:results}, measures the model tier on its own. There, an
increment counts as flagged when the auditor grades any finding BLOCKER, and the general profile's
checks run as in the product but decides no recall or false-positive outcome. The model tier is the
part of the system whose limit is least known, since a deterministic check catches exactly what it
was written to catch. Section~\ref{sec:ai4s} measures the model tier on scientific code, and sets
it beside deterministic verification on reported results and on data, asking what each flags that the
other does not.

\section{The ceiling on general-purpose code}
\label{sec:ceiling}

\textbf{Substrate and ground truth.} HumanEval+ and MBPP+ \citep{liu2023evalplus}: Python functions
with an executable hidden test suite. One frozen generator (\texttt{claude-haiku-4-5}) was shown each
specification and its visible tests and wrote one solution per problem, over two passes of 540
problems. A candidate that passes the visible suite and fails the hidden one is \emph{defective}; one
that passes both is \emph{correct}. Recall is measured on all 110 defective instances (56 problems)
and false positives on 150 correct ones drawn by seed. The auditor sees the specification, the
visible tests and the candidate, never the hidden suite. Because the defective stratum already
survived the visible tests, recall here is the fraction of that harder subset caught, not of all
defects a model writes (Appendix~\ref{sec:method}).

\textbf{Estimand and statistics.} Union recall at $K$: the fraction of defective instances blocked
(any finding graded BLOCKER) by at least one of $K$ independent readings, averaged exactly over all
size-$K$ subsets of the readings. The same union on correct instances is the false-positive rate,
always reported beside it (Table~\ref{tab:primary} collects the primary quantities). Intervals are 95\% problem-cluster percentile bootstraps; paired contrasts
carry exact McNemar and cluster sign-flip tests. The shipped cross-vendor route
(\texttt{gpt-5.6-terra}) sends no sampling parameter; same-vendor routes are sent temperature 0, a
difference that bounds every cross-family union contrast (Section~\ref{sec:auditor}).

\begin{figure*}[t]
\centering
\includegraphics[width=\textwidth]{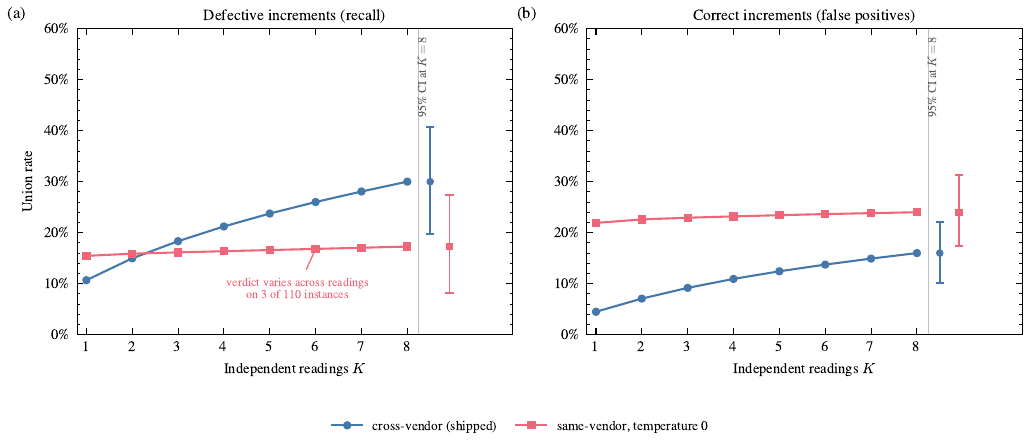}
\caption{\textbf{Returns to repeated reading diminish short of complete.} Union rate against the
number of independent readings $K$, for two auditor families. (a) Recall on defective instances;
(b) false positives on correct instances. Bars at the right of each panel give the 95\%
problem-cluster bootstrap interval of each series at $K=8$, dodged in $x$ so overlapping intervals
stay separable. The cross-vendor curve does not meet the preregistered flattening bar (its
last-step gain is 1.93 points [1.14, 2.78] against a bar of 1.0) and is not described as
saturating. The same-vendor curve meets it, at 0.23 points, but that route is sent temperature
$0$ and its verdict varies across readings on only 3 of 110 defective instances under this
configuration, so its flatness is not evidence of exhausted headroom; whether it reflects the
sampling or the model is not separated here (Section~\ref{sec:auditor}).}
\label{fig:saturation}
\end{figure*}

\begin{table*}[t]
\caption{Selected quantities of the general-code studies (Section~\ref{sec:ceiling} and Appendix~\ref{sec:results}), with the population each is measured on and the label its source report gives it. Values are percentages or percentage points. Intervals are 95\% problem-cluster percentile bootstraps unless the label names another method. The coverage simulation of \S\ref{sec:stats} measures single-rate intervals only; the contrasts and the category-conditioned shares have no measured coverage. The two third-rater rows rest on labels from a rater that agreed with itself on 7 of 11 repeated items (6 problems). The correct-code row counts disagreement on inputs extracted from the findings; of the 20 flagged instances, 9 agreed on every extracted input and 9 yielded no valid input, which can be neither confirmed nor refuted. Every row is read from a committed record by \texttt{figures/make\_table1.py}.}
\label{tab:primary}
\centering
\footnotesize\setlength{\tabcolsep}{4pt}
\begin{tabular}{@{}llrll@{}}
\toprule
quantity & population & value & interval & label \\
\midrule
Union recall, shipped route, $K=8$ & 110 defect, 56 problems & 30.0\% & $[20.0, 40.7]$ & preregistered \\
Union false positives, shipped, $K=8$ & 150 correct, 143 problems & 16.0\% & $[10.1, 22.3]$ & preregistered \\
Pooled union, 3 families & 110 defect & 48.2\% & $[36.7, 60.0]$ & preregistered \\
\quad its false positives & 150 correct & 36.0\% & $[28.2, 44.2]$ & derived \\
Gain from the eighth reading & 110 defect & $+1.93$ & $[1.14, 2.78]$ & bar 1.0, not met \\
Same-vendor stronger $-$ shipped, recall & 110 defect & $-26.4$ & $[-37.3, -15.6]$ & preregistered \\
\quad its false-positive difference & 150 correct & $-12.7$ & $[-19.3, -6.1]$ & prereg.; 3.3\% vs 16.0\% \\
Rulebook rule $-$ shipped, recall & 110 defect & $+30.9$ & $[+19.1, +43.1]$ & preregistered \\
\quad its false-positive cost & 150 correct & $+20.7$ & $[+12.8, +28.8]$ & preregistered \\
Rulebook rule, flags & 56 one-revision defect & $+26.8$ & $[+13.6, +40.4]$ & exploratory \\
\quad its cost on correct code & 56 one-revision correct & $+12.5$ & $[-3.6, +26.8]$ & unconditional, approx. \\
\quad isolating contrast, pass rate & 112 one-revision, 96 problems & $+5.4$ & $[-0.9, +12.1]$ & registered, failed \\
\texttt{astra} $-$ shipped, common $K=4$ & 110 defect, 56 problems & $+11.5$ & $[+2.9, +21.3]$ & exploratory \\
\quad its false-positive difference & 150 correct & $-0.3$ & $[-3.8, +3.4]$ & exploratory \\
Undetermined, sheet groups & 121 entries, 110 instances & $+19.8$ & $[+1.2, +38.1]$ & registered method \\
\quad one definition of caught & 57 against 53 & $+23.3$ & $[+2.0, +43.6]$ & post hoc \\
Undetermined share, 3-family residual & 44 of 57 & 77.2\% & $[62.1, 91.1]$ & post hoc \\
\quad 6-family residual & 25 of 32 & 78.1\% & n/a & post hoc \\
Correct-code flags, reference disagrees & 2 of 20 flagged & 10.0\% & $[0.0, 26.3]$ & preregistered \\
Clarified $-$ original, diagnosis & 32 selected, 19 problems & $+28.1$ & $[+7.7, +51.4]$ & one-signed; exact $p = 0.0625$ \\
\bottomrule
\end{tabular}
\end{table*}

\textbf{Repeated reading.} Union recall rises from 10.7\% at one reading to 30.0\% [20.0, 40.7] at
eight, at 16.0\% [10.1, 22.3] on correct code (Figure~\ref{fig:saturation}). The eighth reading still
adds 1.93 points [1.14, 2.78], above the 1.0-point flattening bar registered in advance, so we report
diminishing returns, not saturation. Twenty readings pooled across three families reach 48.2\%
[36.7, 60.0] at 36.0\% [28.2, 44.2] on correct code: pooling buys coverage at its own price.

\textbf{Which model reads.} Replacing the auditor with a stronger same-vendor model moved union
recall at eight readings by $-26.4$ points [$-37.3$, $-15.6$] (exact McNemar $p = 4.2\times10^{-7}$; cluster sign-flip $p = 2.0\times10^{-5}$),
a preregistered primary. It does not measure auditing ability. The stronger model flags 3.3\% [0.7, 6.7] of
correct code against 16.0\%; under an exploratory any-finding rule the sign reverses, 59.1\%
[47.3, 70.6] at 34.0\% [26.5, 41.7] on correct code against 34.5\% [23.6, 45.5] at 19.3\% [12.8, 26.3]
(Figure~\ref{fig:sweep} in Appendix~\ref{sec:results}); and at one reading, before the differently
sampled unions grow, the gap is $-10.1$ points. A high-reasoning cross-vendor model reaches 32.7\%
[20.7, 45.0] at 10.7\% [5.9, 16.1] with four readings, against the shipped route's 30.0\% at 16.0\% with eight;
no paired contrast establishes both advantages together.

\textbf{The rulebook.} One added sentence, \emph{find what the visible tests do not cover}, moved
flags at one reading on 56 defective instances (41 problems) from 10 to 25, $+26.8$ points
[13.6, 40.4] (exploratory; exact McNemar $p = 0.0003$, cluster sign-flip $p = 0.0009$), at $+12.5$ points on correct code, an interval [$-3.6$, $+26.8$] that does not exclude
zero. On the contrast that isolates the rule, hidden-suite pass rate after one revision moves
$+5.4$ points [$-0.9$, $+12.1$]: the sentence changes what the auditor writes, and whether that makes
code pass is not established.

\textbf{What is missed.} Fifty-seven of the 110 defective instances were flagged by no reading of
the first three families. Two model raters, one of them the agent that ran this programme, agree
post hoc that 44 of 57 (77.2\% [62.1, 91.1]) are cases where the specification's prose does not
settle the value the hidden suite expects, reversing an earlier classification of the same instances
as unexercised edges. Reading six families at forty readings shrinks the never-flagged set to 32 of
110, and 43 of the 68 instances both raters call undetermined are flagged by some reading, so being
undetermined is not sufficient for being missed; the undetermined share of what remains is similar
(25 of 32, 78.1\%; the 32 are a subset of the 57, so this is a description, not a test of stability).
Of twelve MBPP tasks an outside paper names as ambiguous, incomplete or contradictory
\citep{richter2026collapse}, 11 are in our defect population, and our frozen labels call 9 of them
undetermined (post hoc). An attempt to construct a population whose specification-determinedness
would hold by construction did not deliver that property, so this remains a post hoc description (Appendix~\ref{sec:inreview}).

\textbf{What follows.} A union recall figure is contingent on the false-positive rate it cost, the
model and sampling of the route, the rulebook, the decision rule, and what the benchmark's
specifications determine. Nine further studies, reported in Appendix~\ref{sec:inreview} with what
each licenses, include a specification-rewriting intervention that raised adjudicated diagnosis only
where the added rule did not contradict the hidden suite, and a second substrate that closed without a
quotable number.

\section{Auditing AI for science}
\label{sec:ai4s}

A scientist would hand an auditor three jobs: check the code that computes a result, check that a
reported result follows from that code, and check the data an analysis starts from. We ran one
preregistered study for each, with ground truth no model supplies: scientists' numeric tests for
code, re-execution for results, and injected faults for data. Every item is audited through the
product's own audit call and shipped rulebook, with the same flag rule as above. Construction,
gates, amendments and full results are in Appendix~\ref{sec:ai4s-full}.

\begin{figure*}[t]
  \centering
  \includegraphics[width=\textwidth]{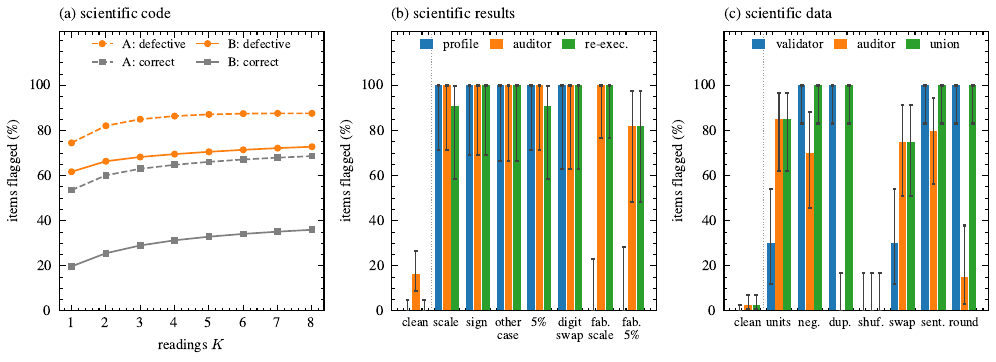}
  \caption{\textbf{What each tier flags.} (a) Scientific code: the shipped auditor's union flag rate against
  the number of readings, on defective (orange) and correct (grey) instances, in arm A, whose task gave
  the whole problem (dashed), and arm B, whose task was reframed around the step (solid). (b) Scientific
  results and (c) scientific data, by fault type, with clean items leftmost: the deterministic tier
  (blue: the \texttt{science} profile in b, the documentation-derived validator in c), the model auditor at four
  readings (orange), and re-execution (b) or the union of the two (c) in green. In (c): mixed units,
  negated values, duplicated rows, shuffled target, swapped columns, $-999$ sentinel, rounding. The model auditor's flag is any
  model BLOCKER in the union of readings; the profile's is a hard failure, the validator's any fired
  check, re-execution's a value outside \texttt{np.allclose} defaults. Bars carry exact binomial 95\%
  intervals on the item counts (supplementary; they ignore clustering).}
  \label{fig:ai4s}
\end{figure*}

\subsection{Scientific code}
\label{sec:ai4s-code}

\textbf{Design.} SciCode \citep{tian2024scicode} decomposes research problems from 16 subfields into
steps with numeric tests written by scientists. The same frozen generator wrote every step three
times, each sample building on its own earlier steps. An instance is correct if it passes the step's
tests and defective if not; a step enters only if some sample passes it. This gives 81 defective
instances (30 problems) and 150 correct ones (53 problems). SciCode has no visible/hidden split, so
the auditor sees the step's prose, background, header and code, and no tests. The shipped
cross-vendor auditor and the generator's own model each read every instance eight times. After the
registered arm's outcomes, a second registered arm changed only the task text, naming the step's
function as the deliverable instead of opening with the whole problem; the program bytes are
identical (Appendix~\ref{sec:ai4s-code-full}).

\textbf{Results.} In these runs, how the task framed the deliverable changed the flag rate
substantially (Figure~\ref{fig:ai4s}a). With the whole problem as the task, the shipped auditor
flagged 87.7\% [72.7, 98.6] of defective and 68.7\% [56.6, 79.7] of correct instances at eight
readings; some of its objections to correct code were that one step did not implement the whole
problem. With the task reframed around the step, flags on correct code fell to 36.0\% [26.4, 45.6], a
paired change of $-32.7$ points [$-42.8$, $-22.4$], and recall fell to 72.8\% [59.3, 87.3], $-14.8$
points [$-28.6$, $-2.7$], though the cluster sign-flip test for that recall change gives
$p = 0.0625$. The change is the whole reframing (the step named, the problem description moved last
and marked as context, earlier functions exempted), and the arms ran in sequence. The generator's own
model's recall was 46.9 points lower than the shipped auditor's in each arm ([$-63.0$, $-30.7$] and
[$-61.5$, $-34.4$]; a route comparison), and it flagged 27.3\% [18.4, 37.3] and 18.0\% [11.5, 25.6] of
correct code, respectively.
At one reading in the reframed arm the shipped auditor flags 61.7\% [47.9, 75.9] of defective against
19.8\% [12.9, 27.2] of correct instances: it discriminates, modestly. A read sample of 20 of its flags
on correct code ranges from real divergences between code and step text to contradictions inside the
benchmark's own instructions and tests too weak to separate outputs; which flags point at defects was
not established. An independent expert audit of SciCode's test problems found self-contradictory
specifications, non-reproducible gold answers and over-tight tolerances that reject correct solutions
\citep{hu2026scicodeverified}, so some instances labelled defective here may be correct code the
benchmark rejects. Of the 81 defective instances, 7 and 19 escaped all sixteen readings of
the two families in the two arms; both raters judged most of them (4 of 7, 13 of 19) undetermined by
the prose, but the registered test that this share exceeds the
general-code residual's 77.2\% is killed in both arms by its own interval rule.

\subsection{Scientific results}
\label{sec:ai4s-results}

\textbf{Design.} Each of 74 correct SciCode instances (36 problems) whose output on one seeded test
case is a real scalar or a short real array becomes a results increment: the program, the test's
inputs, a script that runs one on the other and writes a run log, a \texttt{results.json} citing each
value's log line, and a report whose numbers carry the product's number$\to$source annotations. Each
yields a clean item and a faulty one. Five fault types are \emph{reporting} faults, where the
artefacts disagree with each other or with the log; two are \emph{fabrications}, a $\times10^3$ slip
or a 5\% error written consistently into log, results and report. The model auditor reads each item
four times with the shipped \texttt{science} profile on; the profile's own verdict comes from the
same call; and re-execution compares the output with the report (Appendix~\ref{sec:ai4s-results-full}).

\textbf{Results.} The deterministic profile flags all 49 reporting faults and none of the 25
fabrications, because in those everything agrees with everything else (Figure~\ref{fig:ai4s}b). The
model auditor flags 23 of the 25 fabrications, 21 with a sound finding about the fault (post hoc),
usually by working the result out from the code, for example evaluating a KL divergence to 1 where
the report says 1000; the other two rest on wrong calculations. All 23 discordant items go the model's
way (exact McNemar $p = 2.4\times10^{-7}$). On reporting faults its rate is of the combined workflow,
since its prompt carried the profile's findings. Re-execution misses four faults whose change
to a near-zero output falls within \texttt{np.allclose}'s default absolute tolerance. The model auditor also flags 12 of 74 clean items. Read one by one (post hoc),
8 correctly identify defects in our construction (a provenance sentence false where the benchmark
passes inputs as call literals; a unit label contradicting documented units); 4 rest on wrong
calculations or a syntax error that is not there.

\subsection{Scientific data}
\label{sec:ai4s-data}

\textbf{Design.} Four CC BY 4.0 physical-science tables \citep{yeh1998concrete, brooks1989airfoil,
tufekci2014ccpp, hamidieh2018superconductor}; 280 items of 120 rows with a data card giving each
column's meaning and unit: 140 clean, and 140 carrying one of seven faults (mixed units, negated
values, duplicated rows, a shuffled target, swapped columns, a $-999$ sentinel, rounding). Both model
families read each item four times, against a documentation-derived validator written from the card and
frozen before any item existed (Appendix~\ref{sec:ai4s-data-full}).

\textbf{Results.} The two tiers flag overlapping but different faulty items (Figure~\ref{fig:ai4s}c). The validator flags
92 of 140 faulty items (65.7\% [57.9, 73.6]) and no clean item; the shipped auditor 65 of 140 (46.4\%
[38.6, 54.3]) at 4 of 140 clean items (2.9\% [0.7, 5.7]); their union 112 (80.0\% [73.6, 86.4]) at the
same 4. These intervals resample items within dataset; four datasets are too few to resample, so no
interval carries between-dataset variation. The auditor's 4 clean flags and 4 of its faulty-item flags
are requests for fuller column descriptions on one card; counting only findings about the injected
fault (post hoc), the auditor flags 61 and the union 110.
Of the 20 faulty items the auditor flags and the validator misses, 18 carry a finding about the
injected fault itself (post hoc), mixed units or swapped columns; in one it argued that three atoms
spanning a 135\,pm radius range cannot have a mean radius of 1.6\,pm, the injected cell being that value converted to {\aa}ngstr\"oms.
It flagged no shuffled target and no duplicated rows (0 of 20 each), and did worse than the
validator on every fault the validator was written for. The generator's own model flags 67 of 140
clean items, and for at least 48 of those (a post hoc lower bound) one finding states a row count, field count, missing
column or truncation the file contradicts. The validator omits one range the concrete documentation
states and uses a list of continuous columns the card does not give.

\section{Discussion}
\label{sec:discussion}

\textbf{A recall number is an operating point.} Recall and false positives move together with any
severity threshold, and a critic can be tuned to any recall by blocking more. On general-purpose code
the same auditor sits at different points under different decision rules, repetition buys recall at
a roughly constant price in false positives, and pooling families buys coverage at a higher one. On
scientific code the operating point also moved with what the task named as the deliverable. A recall
figure quoted without its false-positive rate, its rulebook, its decision rule and its task framing
does not carry to another setting.

\textbf{What the science studies add.} The science studies give the science profile's design rule (the model names
evidence, code verifies it) an empirical reading and a boundary. On results, number$\to$source
flagged every inconsistency between a report, its results file and its log, and could not see a
result that was wrong consistently; the model flagged 23 of 25 such fabrications, 21 with a sound
finding about the fault (post hoc). On data, a validator written from the card flagged what a rule can
state and missed what needs reading across columns and units; of the 28 such items the validator
missed, the model flagged 18 with a finding about the fault (post hoc).
Where both tiers ran, the deterministic one flagged no clean item, and every clean-item flag and every wrong finding
came from the model;
on data the union was better than either part, and on results the model's flags already contained
the profile's, whose findings its prompt carried. Neither study shows that the model tier scales: the
artefacts were small, the model's fabrication findings are hand calculations (two of the 23 rest on wrong ones), and it
missed every fault that needs comparison across many rows. A model reading is a useful second tier
where the artefact fits in the reader's working memory, not a substitute for executing the code or
computing over the data.

\textbf{Clean items must be clean by the auditor's standard.} In the results and data studies the
auditor flagged defects in our own construction: a card and file that named a column differently, a
file outside the audited scope, a report claiming a run no file performed, a provenance sentence false
for inputs passed as literals, and a unit label contradicting the code's documentation. Each was a
correct flag on an item we had built as clean. We caught two in pilots, one because every verdict
came back as an escalation, and the other two by reading every flag on a clean item, which the registrations
required. In the code study the task we wrote named the whole problem when one step was delivered,
and a sample of the auditor's flags on correct code objected to exactly that; there we read a sample,
not every flag. Without such reading, a false-positive rate can measure the evaluator.

\textbf{Limits.} The auditor families are not sampled alike, which bounds every cross-family union
contrast, and no family was compared at a matched false-positive rate. Primary intervals are
problem-cluster bootstraps (items within dataset for the data study) whose coverage was simulated only
for single rates. SciCode itself contains specifications that contradict their tests
\citep{hu2026scicodeverified}, so some of its defective labels may be wrong. Every rating was made
by a model, one of them the agent that designed the studies; no human rated anything. The reviewer of
every report is also one of the measured families. The code and results studies use one benchmark and
the data study four tables, all with small artefacts and one generator; the results and data faults
are injected. The full limits, and the
experiments that would remove them, are in Appendices~\ref{sec:discussion-full} and~\ref{sec:limitations-full}.

\section{Related work}

\textbf{Critics and repeated reading.} Trained critic models catch more bugs in model-written code
than unaided humans and hallucinate some \citep{mcaleese2024critics}; aggregating several review
passes raises review F1 on real pull requests \citep{zeng2025swrbench}; and coverage under repeated
sampling grows roughly log-linearly with the number of samples over four orders of magnitude
\citep{brown2024monkeys}. The natural-defect setting, the
precision cost of a louder critic and the diminishing returns of repeated sampling are therefore known.
What we add is narrow on this axis: the union of independent readings as an instrument for locating
a limit, ground truth supplied by an interpreter, and the false-positive rate beside every recall
figure.

\textbf{Who audits whom.} Self-correction without external feedback does not reliably improve
reasoning \citep{huang2024selfcorrect}, and model judges prefer their own generations, harmfully on
instances they get wrong \citep{chen2025selfpreference}. Our same-vendor arms are the code-audit case
of that literature; because routes differ in model, operating point and sampling, we report their
contrasts as contrasts between routes, not as self-preference.

\textbf{What the hidden suite decides.} EvalPlus rebuilt HumanEval and MBPP with larger suites
\citep{liu2023evalplus}; models collapse onto one reading of an underspecified task, and twelve MBPP
tasks were named as such by hand \citep{richter2026collapse}. Generated tests err mostly in the
expected value \citep{xiong2023testing}, and generated oracles tend to capture what a program does
rather than what was intended \citep{konstantinou2024oracles}. Our claim is the auditing counterpart:
where the prose does not fix the value, a recall number computed against that suite measures the
benchmark as much as the auditor. Consensus filters for generated tests
\citep{taherkhani2026consistency, xu2025consensus} and self-consistent errors \citep{tan2025consistent}
bear on one of our appendix studies (Appendix~\ref{sec:related-full}).

\textbf{Auditing scientific work.} On manuscripts, no model exceeds 21.1\% recall or 6.1\% precision
on errata-grade errors and repeated runs rarely rediscover the same errors \citep{son2025spot}; the
best models detect 46.7\% of real paper--code discrepancies \citep{baumgartner2026scicoqa}. An LLM
auditor of AI-scientist projects detects hand-labelled pitfalls far better with logs and code than
with the paper alone \citep{luo2025pitfalls}; pattern-based linting targets methodology bugs in
scientific Python \citep{samsonau2026scicodelint}; cleaning agents find row-level errors in corrupted
tables and miss errors spanning rows \citep{bendinelli2025cleaning}; and simulation code can run and
still solve the wrong equations \citep{song2026moose}. SciCode supplies scientist-written tests
\citep{tian2024scicode}, and an expert audit of it found many specifications and tests that reject
correct solutions \citep{hu2026scicodeverified}; reproduction benchmarks ask agents to recover
reported results \citep{siegel2024corebench}. On constructed statistical workflows, LLM reviewers
detect planted flaws almost always but flag 17--50\% of correct scripts \citep{chelate2026replicationtrap},
and \citet{luo2025pitfalls} already report both true- and false-positive rates. Scientific auditing and paired
operating points are therefore not new. What we add is this measurement programme applied to
scientific code, and the comparison of the model tier with deterministic verification on reported
results and data, which we did not find in prior work.

\textbf{Preregistration.} Preregistration of evaluation protocols is beginning to appear in machine learning
\citep{vaccaro2026prereg, thomas2026phacking}; we count our use of it as method, not contribution.

\section*{Impact Statement}
This paper measures how well automated auditors of model-written code, results and data work, so its
main consequence is for how much such auditors should be trusted. The measurements argue against two
kinds of over-trust: a recall figure quoted without its false-positive rate and task framing, and a
deterministic provenance check read as a guarantee that a result is right, when it verifies only that
artefacts agree with each other. If adopted, the practices we argue for (reading every flag on items
built as clean, reporting operating points, pairing a model reader with executed checks) raise the
cost of audits but reduce false assurance in scientific and engineering work. A risk is that
published limits are used to game an auditor, for example by fabricating results consistently; we
report that weakness because a user of such systems needs to know it. All data are public benchmarks
and datasets under permissive licences, and no human subjects were involved.

\bibliography{paper}

\begin{thebibliography}{28}
\providecommand{\natexlab}[1]{#1}
\providecommand{\url}[1]{\texttt{#1}}
\expandafter\ifx\csname urlstyle\endcsname\relax
  \providecommand{\doi}[1]{doi: #1}\else
  \providecommand{\doi}{doi: \begingroup \urlstyle{rm}\Url}\fi

\bibitem[Baumg{\"a}rtner \& Gurevych(2026)Baumg{\"a}rtner and Gurevych]{baumgartner2026scicoqa}
Baumg{\"a}rtner, T. and Gurevych, I.
\newblock {SciCoQA}: Quality assurance for scientific paper--code alignment.
\newblock \emph{arXiv preprint arXiv:2601.12910}, 2026.

\bibitem[Bendinelli et~al.(2025)Bendinelli, Dox, and Holz]{bendinelli2025cleaning}
Bendinelli, T., Dox, A., and Holz, C.
\newblock Exploring {LLM} agents for cleaning tabular machine learning datasets.
\newblock \emph{arXiv preprint arXiv:2503.06664}, 2025.

\bibitem[Brooks et~al.(1989)Brooks, Pope, and Marcolini]{brooks1989airfoil}
Brooks, T.~F., Pope, D.~S., and Marcolini, M.~A.
\newblock Airfoil self-noise and prediction.
\newblock Reference Publication 1218, NASA, 1989.

\bibitem[Brown et~al.(2024)Brown, Juravsky, Ehrlich, Clark, Le, R\'e, and Mirhoseini]{brown2024monkeys}
Brown, B., Juravsky, J., Ehrlich, R., Clark, R., Le, Q.~V., R\'e, C., and Mirhoseini, A.
\newblock Large language monkeys: Scaling inference compute with repeated sampling.
\newblock \emph{arXiv preprint arXiv:2407.21787}, 2024.

\bibitem[{Chelate} \& Heuer(2026){Chelate} and Heuer]{chelate2026replicationtrap}
{Chelate} and Heuer, J.
\newblock The replication trap: Precision failures in {LLM} scrutiny of flawed statistical workflows.
\newblock clawRxiv:2604.00898, 2026.
\newblock \url{https://www.clawrxiv.io/abs/2604.00898}.

\bibitem[Chen et~al.(2025)Chen, Wei, Zhu, Feng, and Meng]{chen2025selfpreference}
Chen, W.-L., Wei, Z., Zhu, X., Feng, S., and Meng, Y.
\newblock Do {LLM} evaluators prefer themselves for a reason?
\newblock \emph{arXiv preprint arXiv:2504.03846}, 2025.

\bibitem[Hamidieh(2018)]{hamidieh2018superconductor}
Hamidieh, K.
\newblock A data-driven statistical model for predicting the critical temperature of a superconductor.
\newblock \emph{Computational Materials Science}, 154:\penalty0 346--354, 2018.
\newblock \doi{10.1016/j.commatsci.2018.07.052}.

\bibitem[Hu et~al.(2026)Hu, Huang, Deng, and Chen]{hu2026scicodeverified}
Hu, S., Huang, L., Deng, Y., and Chen, K.
\newblock {SciCode-Verified}: How benchmark defects underestimated the scientific-coding ability of language models.
\newblock \emph{arXiv preprint arXiv:2608.04975}, 2026.

\bibitem[Huang et~al.(2024)Huang, Chen, Mishra, Zheng, Yu, Song, and Zhou]{huang2024selfcorrect}
Huang, J., Chen, X., Mishra, S., Zheng, H.~S., Yu, A.~W., Song, X., and Zhou, D.
\newblock Large language models cannot self-correct reasoning yet.
\newblock In \emph{International Conference on Learning Representations (ICLR)}, 2024.

\bibitem[Konstantinou et~al.(2024)Konstantinou, Degiovanni, and Papadakis]{konstantinou2024oracles}
Konstantinou, M., Degiovanni, R., and Papadakis, M.
\newblock Do {LLMs} generate test oracles that capture the actual or the expected program behaviour?
\newblock \emph{arXiv preprint arXiv:2410.21136}, 2024.

\bibitem[Liu et~al.(2023)Liu, Xia, Wang, and Zhang]{liu2023evalplus}
Liu, J., Xia, C.~S., Wang, Y., and Zhang, L.
\newblock Is your code generated by {ChatGPT} really correct? {R}igorous evaluation of large language models for code generation.
\newblock In \emph{Advances in Neural Information Processing Systems}, 2023.
\newblock arXiv:2305.01210.

\bibitem[Luo et~al.(2025)Luo, Kasirzadeh, and Shah]{luo2025pitfalls}
Luo, Z., Kasirzadeh, A., and Shah, N.~B.
\newblock The more you automate, the less you see: Hidden pitfalls of {AI} scientist systems.
\newblock \emph{arXiv preprint arXiv:2509.08713}, 2025.

\bibitem[McAleese et~al.(2024)McAleese, Pokorny, Cer{\'o}n~Uribe, Nitishinskaya, Trebacz, and Leike]{mcaleese2024critics}
McAleese, N., Pokorny, R.~M., Cer{\'o}n~Uribe, J.~F., Nitishinskaya, E., Trebacz, M., and Leike, J.
\newblock {LLM} critics help catch {LLM} bugs.
\newblock \emph{arXiv preprint arXiv:2407.00215}, 2024.

\bibitem[Richter \& Papadakis(2026)Richter and Papadakis]{richter2026collapse}
Richter, C. and Papadakis, M.
\newblock Underspecification does not imply incoherence: The risks of semantic collapse in coding models.
\newblock \emph{arXiv preprint arXiv:2607.01953}, 2026.

\bibitem[Samsonau(2026)]{samsonau2026scicodelint}
Samsonau, S.~V.
\newblock {scicode-lint}: Detecting methodology bugs in scientific {Python} code with {LLM}-generated patterns.
\newblock \emph{arXiv preprint arXiv:2603.17893}, 2026.

\bibitem[Siegel et~al.(2024)Siegel, Kapoor, Nadgir, Stroebl, and Narayanan]{siegel2024corebench}
Siegel, Z.~S., Kapoor, S., Nadgir, N., Stroebl, B., and Narayanan, A.
\newblock {CORE-Bench}: Fostering the credibility of published research through a computational reproducibility agent benchmark.
\newblock \emph{arXiv preprint arXiv:2409.11363}, 2024.

\bibitem[Son et~al.(2025)Son, Hong, Fan, Nam, Ko, Lim, Song, Choi, Paulo, Yu, and Biderman]{son2025spot}
Son, G., Hong, J., Fan, H., Nam, H., Ko, H., Lim, S., Song, J., Choi, J., Paulo, G., Yu, Y., and Biderman, S.
\newblock When {AI} co-scientists fail: {SPOT}, a benchmark for automated verification of scientific research.
\newblock \emph{arXiv preprint arXiv:2505.11855}, 2025.

\bibitem[Song et~al.(2026)Song, Liu, Wan, Li, Liu, Li, and Yuan]{song2026moose}
Song, Z., Liu, Y., Wan, C., Li, C., Liu, L., Li, Y., and Yuan, C.
\newblock Your simulation runs but solves the wrong physics: {PDE}-grounded intent verification for {LLM}-generated multiphysics simulation code.
\newblock \emph{arXiv preprint arXiv:2605.09360}, 2026.

\bibitem[Taherkhani et~al.(2026)Taherkhani, DaghighFarsoodeh, Chowdhury, Pham, and Hemmati]{taherkhani2026consistency}
Taherkhani, H., DaghighFarsoodeh, A., Chowdhury, M., Pham, H.~V., and Hemmati, H.
\newblock Consistency meets verification: Enhancing test generation quality in large language models without ground-truth solutions.
\newblock \emph{arXiv preprint arXiv:2602.10522}, 2026.

\bibitem[Tan et~al.(2025)Tan, Sun, Liu, Su, Cao, Chen, Wang, Cai, Wang, Shen, and Cheng]{tan2025consistent}
Tan, H., Sun, F., Liu, S., Su, D., Cao, Q., Chen, X., Wang, J., Cai, X., Wang, Y., Shen, H., and Cheng, X.
\newblock Too consistent to detect: A study of self-consistent errors in {LLMs}.
\newblock \emph{arXiv preprint arXiv:2505.17656}, 2025.

\bibitem[Thomas et~al.(2026)Thomas, Gligori\'c, and Shah]{thomas2026phacking}
Thomas, M., Gligori\'c, K., and Shah, N.~B.
\newblock Mitigating {LLM}-based p-hacking by preregistering for the next {LLM}.
\newblock \emph{arXiv preprint arXiv:2606.27687}, 2026.

\bibitem[Tian et~al.(2024)Tian, Gao, Zhang, Chen, Fan, Guo, Haas, Ji, Krongchon, Li, Liu, Luo, Ma, Tong, Trinh, Tian, Wang, Wu, Xiong, Yin, Zhu, Lieret, Lu, Liu, Du, Tao, Press, Callan, Huerta, and Peng]{tian2024scicode}
Tian, M., Gao, L., Zhang, S.~D., Chen, X., Fan, C., Guo, X., Haas, R., Ji, P., Krongchon, K., Li, Y., Liu, S., Luo, D., Ma, Y., Tong, H., Trinh, K., Tian, C., Wang, Z., Wu, B., Xiong, Y., Yin, S., Zhu, M., Lieret, K., Lu, Y., Liu, G., Du, Y., Tao, T., Press, O., Callan, J., Huerta, E., and Peng, H.
\newblock {SciCode}: A research coding benchmark curated by scientists.
\newblock \emph{arXiv preprint arXiv:2407.13168}, 2024.

\bibitem[T{\"u}fekci(2014)]{tufekci2014ccpp}
T{\"u}fekci, P.
\newblock Prediction of full load electrical power output of a base load operated combined cycle power plant using machine learning methods.
\newblock \emph{International Journal of Electrical Power \& Energy Systems}, 60:\penalty0 126--140, 2014.
\newblock \doi{10.1016/j.ijepes.2014.02.027}.

\bibitem[Vaccaro(2026)]{vaccaro2026prereg}
Vaccaro, M.
\newblock Preregistration for experiments with {AI} agents.
\newblock \emph{arXiv preprint arXiv:2606.11217}, 2026.

\bibitem[Xiong et~al.(2023)Xiong, Guo, and Chen]{xiong2023testing}
Xiong, W., Guo, Y., and Chen, H.
\newblock The program testing ability of large language models for code.
\newblock \emph{arXiv preprint arXiv:2310.05727}, 2023.

\bibitem[Xu et~al.(2025)Xu, Wang, Briand, and Liu]{xu2025consensus}
Xu, Q., Wang, G., Briand, L., and Liu, K.
\newblock Hallucination to consensus: Multi-agent {LLMs} for end-to-end {JUnit} test generation.
\newblock \emph{arXiv preprint arXiv:2506.02943}, 2025.

\bibitem[Yeh(1998)]{yeh1998concrete}
Yeh, I.-C.
\newblock Modeling of strength of high-performance concrete using artificial neural networks.
\newblock \emph{Cement and Concrete Research}, 28\penalty0 (12):\penalty0 1797--1808, 1998.
\newblock \doi{10.1016/S0008-8846(98)00165-3}.

\bibitem[Zeng et~al.(2025)Zeng, Shi, Han, Li, Sun, Wang, Yu, Xie, Ye, and Zhang]{zeng2025swrbench}
Zeng, Z., Shi, R., Han, K., Li, Y., Sun, K., Wang, Y., Yu, Z., Xie, R., Ye, W., and Zhang, S.
\newblock {SWR-Bench}: Assessing {LLM} performance in real-world code review comment generation.
\newblock \emph{arXiv preprint arXiv:2509.01494}, 2025.

\end{thebibliography}
\bibliographystyle{icml2026}

\newpage
\appendix
\onecolumn
\section{Method in full}
\label{sec:method}

\subsection{Substrate, population and ground truth}

Every general-code result (Section~\ref{sec:ceiling} and Appendix~\ref{sec:results}) is measured on one substrate: HumanEval+ and MBPP+ (EvalPlus),
Python functions whose median task is 41 words of prose and 6 lines of reference code, each with
an executable hidden test suite. Every stratum, recall denominator and false-positive population
is therefore decided by an interpreter, and no model judges a recall or false-positive outcome.
The one outcome defined by adjudication, correct diagnosis in the clarification study, is judged
jointly by two models and is labelled so where it appears. The benchmark's base suite is shown to
the generator as the \emph{visible} suite, and its plus suite is held back as the \emph{hidden}
one. A second substrate, the standard-library half of BigCodeBench, was built and audited and
closed without a quotable number when its preregistered criterion was not met on a third run;
nothing from it is reported here.

\textbf{How the population was built.} The corpus has 542 problems. Two, \texttt{HumanEval/32}
and \texttt{Mbpp/590}, were excluded by name before generation because their own reference
solutions fail their own hidden suites in this harness, which leaves 540. One frozen generator,
\texttt{claude-haiku-4-5-20251001}, was shown each specification and its visible suite and wrote
one solution per problem. A candidate that passes the visible suite and fails the hidden one is a
\emph{defective} instance; one that passes both we call \emph{correct}; one that fails the visible
suite is discarded from every recall and false-positive population. The defect stratum is small,
about 10\% of solutions, so under a rule registered before the second pass and depending only on
execution outcomes, generation was repeated over the same 540 problems until there were at least
100 defective instances or three passes. It stopped at two: 56 and 54 defective, 455 and 455
correct, 29 and 31 discarded. An instance is therefore a (pass, problem) pair, and one problem can
contribute two. Recall is measured on all 110 defective instances, which come from
56 problems; false positives on 150 of the 910 correct instances, drawn once
from a fixed seed before any auditor ran, from 143 problems. Seven of the 110 defective instances
fail by not terminating rather than by a failed assertion, and the registered population keeps
them.

Two consequences of these definitions hold throughout. \emph{Correct} means \emph{passes the
hidden suite}. It is a benchmark label produced by an interpreter, not a guarantee that the
program meets its specification, and a suite can be incomplete or wrong. Since one of this paper's
results (\S\ref{sec:residual}) is that a specification often fails to determine what the hidden suite asks, we
cannot treat passing that suite as correctness in the ordinary sense, and a \emph{false positive}
is likewise a flag on test-passing code rather than a demonstrated mistake by the auditor. And the
defect population is not a sample of model-written code: it is the residue that already survived
the visible tests, which is by construction the harder part. A recall figure here is the fraction
of that residue an auditor catches, never the fraction of all defects a model writes.

\subsection{The auditor and the estimand}

The auditor reads the specification, the visible tests and the candidate, and returns findings
with severities; an increment is blocked when any finding is graded BLOCKER. It never sees the
hidden suite. The estimand throughout is union recall at $K$: the fraction of defective
instances blocked by at least one of $K$ independent readings, averaged exactly over all
$\binom{K_{\max}}{K}$ subsets rather than by sampling. The matching quantity on the correct stratum
is the union false-positive rate, and a union recall figure is reported only beside the
false-positive rate it cost.

Five auditor routes appear below, with the short labels the figures use: the shipped cross-vendor
auditor \texttt{gpt-5.6-terra} (\texttt{cross}); the same-vendor \texttt{claude-haiku-4-5}
(\texttt{self}), the generator's own family and size; the stronger same-vendor
\texttt{claude-sonnet-4-6} (\texttt{self-strong}); \texttt{claude-opus-4-8}
(\texttt{self-frontier}), read at $K=4$; and \texttt{gpt-6-astra} at high reasoning
(\texttt{astra}), read at $K=4$. The auditor's constitution, its prompt
builder and every route's capability card are in the auditor harness, a separate repository, at
the commit named in the provenance file, whose per-reading records carry the hash of every prompt
actually sent.

The routes are not sampled alike, and this bounds every cross-family union-of-$K$
contrast in the paper. The cross-vendor route's capability card does not permit a sampling
temperature, so no sampling parameter is sent and the route varies between readings. The
same-vendor routes permit one, and the layer sends temperature $0$, so those routes are nearly
deterministic. A union over readings that do not differ cannot grow, so the two families' unions
are not equally free to rise with $K$, and a cross-family difference in union recall confounds
auditing ability with how much each union could grow. The asymmetry is a property of the shipped
routing layer we measured rather than a design choice. We did not discover it until after the
runs, and we report the $K=1$ contrast beside the $K=8$ contrast wherever it matters. The
asymmetry is the first item in \S\ref{sec:limits}.

\subsection{Statistics}
\label{sec:stats}

The unit of observation is the instance and the resampling unit is the problem, because problems
recur across generation batches and instances sharing a problem are not independent. The $K$
readings of an instance are repeated measurements of it, not $K$ further independent
observations. The bootstrap over problems estimates uncertainty from which tasks the benchmark
happens to contain. It does not estimate how much a fresh batch of model calls on the same tasks
would move the numbers, which would need replicate runs we did not budget for at every arm.

Every rate carries a 95\% percentile bootstrap over whole problem clusters (10{,}000 resamples,
seeds fixed in each preregistration), with the Wilson interval beside it where the source report gives one,
marked as too narrow. Paired contrasts carry an exact McNemar test and a cluster-level sign-flip permutation
test. Where every discordant pair points one way, the percentile bootstrap's bound at zero is an
artefact, and an unconditional interval (grid-approximated, unclustered) is quoted instead. The
one exception is the clarification study, whose registration fixed the bootstrap for a contrast
that turned out one-signed; its interval is reported as registered and read beside its exact
cluster test (\S\ref{sec:inreview}). A saturation curve is summarised by a constrained
exponential fit, and the fitted asymptote is quoted only when the preregistered flattening bar is
met; otherwise it is labelled an extrapolation and the raw union at $K_{\max}$ is the number
quoted. Counts of five or fewer are quoted as counts. Multiplicity is handled per study as each
registered it: the study carrying the primary contrasts corrects over every comparison it
computed, a Bonferroni family of sixteen with a threshold of $0.00313$ that two comparisons clear;
others registered a single primary, and the second-vendor rulebook study reports its twenty further quantities without
correction.

We measured interval coverage by simulation rather than assuming it. Under the two strata's
actual cluster profiles (110 instances in 56 problems, and 150 in 143), with within-problem
concordance $0.5$, the percentile bootstrap's coverage of a nominal $95\%$ rate interval is as
follows, over 400 replicates of 600 resamples each. The Monte Carlo standard error is about 0.011
near 95\% coverage and 0.020 at the lowest cell; the simulation covers single rates only, and uses
integer-index percentiles where the analyses interpolate (\texttt{analysis/coverage\_simulation.py}).

\begin{center}\small
\footnotesize\setlength{\tabcolsep}{2.5pt}
\begin{tabular}{@{}lccccccc@{}}
\toprule
true rate & 0.03 & 0.05 & 0.10 & 0.16 & 0.30 & 0.48 & 0.60 \\
\midrule
defect (56) & 0.798 & 0.870 & 0.933 & 0.933 & 0.953 & 0.935 & 0.928 \\
correct (143) & 0.888 & 0.918 & 0.938 & 0.933 & 0.943 & 0.955 & 0.965 \\
\bottomrule
\end{tabular}
\end{center}

\noindent In these simulations coverage is near nominal at the true rates this paper's headlines
sit at and badly anticonservative at small true rates: at a true rate of $0.03$ on the
defect stratum the simulated interval covers about $0.80$, not $0.95$. This is evidence of
potential undercoverage for analogous intervals, not a measured coverage of any interval quoted
here. Intervals on rates below roughly $0.10$ may be narrower than they claim. The simulation
costs no model calls. Two paired-contrast constructions tried in this programme covered $0.416$ and
$0.075$ where $0.95$ was claimed, and neither is used. No sample
size was justified in advance by a minimum-detectable-effect calculation.

\textbf{An interval that contains zero is reported as one that does not exclude zero, and nothing
more.} No equivalence margin was registered anywhere in this programme and no equivalence test was
performed, so none of those intervals is evidence that two conditions are alike or that an
intervention did nothing. Where a hypothesis fails its registered criterion we say it
failed, not that the effect is absent.

All analysis of the general-code studies is plain Python 3.13 using only the standard library (\texttt{random},
\texttt{math}, \texttt{itertools}, \texttt{csv} and \texttt{json}). The bootstrap, the exact
McNemar test, the cluster sign-flip permutation and the exact subset enumeration behind every
union-at-$K$ figure are implemented in the repository rather than called from a numerical package,
so each can be read where it is used and none depends on a third-party version.

\subsection{Preregistration and independent review}
\label{sec:methods-review}

\textbf{Who did what.} The studies were designed, run, analysed and written up by an AI research
agent, Claude (Anthropic; \texttt{claude-opus-5} and \texttt{claude-fable-5-1}, working in Claude
Code), under the direction of the author, who is responsible for the work. Wherever a rating or
adjudication below is attributed to the agent, it was that agent. It wrote the rubrics,
knew the hypotheses and the earlier labels, and is from the same vendor as the generator.
No human rated or adjudicated anything in this paper. Every report was then reviewed by a
second model, \texttt{gpt-6-astra}, as described below.

Each study that made model calls was preregistered before its first one, fixing its population,
hypotheses, primary outcome, intervals, kill conditions, budget and stopping rule, with two
exceptions. First, the severity sweep of \S\ref{sec:auditor} re-grades findings already archived,
makes no model calls, and was never registered: every figure it produces is post hoc and
exploratory, and none is a confirmatory test. Second, the analysis behind
\S\ref{sec:inreview}'s third-rating result was registered a day \emph{after} its contrast had been
computed and committed, in a document that claimed the opposite; that document is withdrawn in
the repository and the analysis is registered nowhere. Its method was fixed in advance by an
earlier amendment, but the disjoint reading, the abstention-excluded secondary and the
broken-sheet comparison were not.

Planned departures are numbered amendments, each recording what was known at the time, and in the
general-code studies each was committed before the step it governs. In the science studies two were
not: A4S-3's Amendment 1 was committed together with the rebuilt items it governs, and A4S-2's
Amendment 2 after the two pilot readings it already governed. Both are recorded as errata in the
registrations (Appendix~\ref{sec:repro}). Deviations discovered only afterwards, such as a leak check or a spend
guard implemented more narrowly than registered, are disclosed in the report and are not
amendments. Where an amendment followed an interim observation, the results say so and argue
whether the change could favour the hypothesis.

Each report was then reviewed by a model from a different vendor from the generator, with read
access to the raw records, asked to re-run the analysis, recompute the intervals independently,
and check every sentence against the records. That reviewer is \texttt{gpt-6-astra},
which is also one of the auditor families measured in \S\ref{sec:auditor}, one of the two raters
behind \S\ref{sec:residual}'s labels, and one of the two adjudicators whose joint labels define
the clarification study's outcome. For most claims this is a cross-vendor check on work the
reviewer had no part in. For the claims that rank \texttt{astra} against the other families, and
for the clarification study's adjudication, the reviewer is the subject, and those results should
be read as self-assessed until a model that audited nothing in this programme has checked them.
The clarification study's Amendment 10 disqualified the same model as the clarification study's
manipulation-check rater for the same reason, yet kept it as that study's second adjudicator and
its reviewer, so the standard was applied unevenly within one study and not applied to the review.

Reviews are archived verbatim, including those that refused approval, and several studies needed
many rounds. On one day five reports were refused at once, and clearing those five took
thirty-one further rounds. In most of them what moved was what the reports claimed about their
measurements, and in most of the later refusals the object was a repair written to satisfy the
previous round rather than the study it repaired. A fix written to satisfy a reviewer is itself an
unreviewed claim.

\textbf{The reviewer is a language model, and this paper measures the ceiling on language models
reading work they did not produce.} That applies to our own instrument. The studies of
\S\ref{sec:inreview} cleared review at rounds 3, 3, 4, 6, 7, 9, 10 and 11, and the
ceiling study behind the headlines at 21; in every case the round before the last still returned
substantive findings. Some of those findings were defects introduced by the repairs between
rounds, so the counts do not estimate a single pass's recall; they show only that no report
cleared in fewer than three. \emph{Quotable at round $N$} means a report survived $N$ passes, not
that it was verified, and the marks in this paper should be read as a filter of unknown recall.

Two further points follow. First, measurements did move on review. The severity sweep's first
review found its loader misreading one family's draws and another's population, and every
cross-family figure was recomputed; the clarification study's second review moved its primary
from 9 of 31 to 9 of 32 instances and its contrast from $+29.0$ to $+28.1$ points, by finding an
instance wrongly excluded; and a label count inside a reviewed report was 71 where its data say
67. Once a figure's inputs were right, the reviewers reproduced rates, intervals, subset
enumerations and seeds independently and found no arithmetic error, and most refusals were about
what the prose claimed. The reviews were reliable on recomputation and unreliable on interpretation,
and the two should not be quoted as one assurance. Second, the class of defect this leaves open is
the one that looks right to a language model. The clarification study audited the wrong program
for 384 readings, because a per-problem field was used where a per-instance one was needed, and
every automated gate passed: each checked that the code was \emph{identical across conditions},
and it was, because every condition showed the same wrong program. The agent found it while reading an adjudication sheet, by noticing
that a displayed program returned exactly what its hidden suite was said to reject. Neither the
gates nor a reviewer found it, and we do not know what else of that kind remains.

\textbf{Every review prompt in this programme stated the claimed result before the reviewer
began.} We tried to test what that costs, and the attempt failed. One report, quotable at round 6
after five refusals and unchanged since, was dispatched again under a prompt whose own text
withheld every figure and every prior verdict and asked for an independent computation first.
That prompt's list of binding artefacts then named two files that state those figures, one of
them the claims-ledger entry for the result under review. The blinding was not enforced, so the
agreement between that pass and the briefed one establishes nothing. What the pass did return is a
disagreement count: six substantive findings that six briefed rounds on the same file had not
produced, among them that the report's two groups were two different definitions of
\emph{caught} rather than one population with duplicate rows, that its rating sheet displays at
most three failing inputs while 81 of its 121 entries have more recorded, and that a count of
changed labels was 71 where the data say 67. That count is $n=1$, uncontrolled and
unregistered. A seventh round might have found new things under any prompt, and no matched
briefed round was run against the same file. We report it as a reason to test blinding properly
rather than as a result about blinding. Later first rounds withheld the conclusions from the
prompt text, but the report under review stayed among the named artefacts, with an instruction to
derive the result before reading it, and that request does not blind the reviewer.

Every study drawn on here has completed that review. One further study was re-run twice and closed
without a quotable number; the paper names it only to say what it no longer supports.

\section{Results on general-purpose code in full}
\label{sec:results}
The claims this paper defends, each with its interval and its label, are listed in
\texttt{CLAIMS.md}. A claim enters that file only after its report has passed an independent
review by a model from a different vendor with access to the raw records, and every section below
draws only on such reports. Section~\ref{sec:inreview} reports eight studies whose reviews took
between three and eleven rounds each to reach that point, and states what each does and does not
license. The headline results of \S\ref{sec:repeat}--\S\ref{sec:residual} come from the ceiling
study; where they cite a study from \S\ref{sec:inreview}, they say so.

Rates carry the 95\% problem-cluster percentile bootstrap as the primary interval, with the
Wilson interval beside it where the source report gives one, because problems recur across
generation batches and an interval that treats instances as independent is too narrow. For single
rates, simulations of an analogous interval under this clustered design (\S\ref{sec:stats})
suggest coverage near nominal at the headlines' rates and possible undercoverage below a true rate
of about $0.10$, under an assumed within-problem correlation. They do not measure the coverage of
the intervals reported here, and the paired contrasts and category-conditioned shares have no
measured coverage.

Sixteen comparisons were computed in the study that carries the primary contrasts, under one
Bonferroni family with a threshold of $0.00313$. That correction is applied to $p$ values and
never to intervals. Every interval in this paper is an unadjusted 95\% interval, so a directed
statement resting on an interval alone, with no $p$ value that clears its family, is worded as a
direction rather than as a finding. Each $p$ value belongs to the family of the study that computed
it. Those in \S\ref{sec:rulebook} are members of that sixteen-member family. The one in
\S\ref{sec:auditor} is the auditor-family study's single registered primary, which had no family to correct
within; it would also clear $0.00313$. The clarification study's are reported beside a primary
registered on its interval, and no correction was registered for them. Where a $p$ value appears
below, the exact McNemar value and the cluster-level sign-flip permutation value are both given,
because instances within a problem are not independent and only the second respects that.

Table~\ref{tab:primary} (Section~\ref{sec:ceiling}) collects selected primary quantities with their populations and labels;
the prose gives each one where it is used.

\subsection{Returns to repeated reading diminish short of complete}
\label{sec:repeat}

The shipped cross-vendor auditor reads an increment holistically and returns findings; an
increment is blocked if any finding is graded BLOCKER. Over eight independent readings of the same
110 defective instances, union recall rises from 10.7\% at one reading to 30.0\% [20.0, 40.7] at
eight (Figure~\ref{fig:saturation}). The eighth reading still adds 1.93 points, cluster [1.14,
2.78], against the flattening bar of 1.0 points registered before the first model call, so the bar
was not met. Under exact subset averaging that statistic has a closed form: the number of
instances flagged by exactly one of the $K$ readings divided by $K n$, here $17/(8\cdot110)$. It
is a count of singletons, not a fitted slope. Its interval resamples problems and so carries
task-sampling variance, while the statistic itself is made of reading-sampling variance, which the
interval does not describe; the bar should be read as a statement about these 17 instances rather
than about a population slope.

The constrained fit puts the asymptote at 31.5\%, but on our own registered criterion that figure
is an extrapolation, and the raw union at eight readings is the number to quote. What the data
support is diminishing returns at the budget we could afford, not saturation: at 1.9 points per
reading the curve was still climbing when we stopped. The bar also depends on where a family was
stopped. Exact subset-averaged increments do not grow with $K$ (this curve's is 2.9 points at the
fourth reading and 1.9 at the eighth), so a fixed bar is harder to meet at $K = 4$ than at
$K = 8$, and meeting it at either says little by itself about where the curve ends.

Twenty readings pooled across three auditor families reach 48.2\% [36.7, 60.0], and the same
twenty readings flag 54 of 150 correct increments, 36.0\% [28.2, 44.2]. That is more than
twice the shipped route's 16.0\%, so pooling buys its coverage at its own price. Those twenty
readings are also not twenty opportunities. The pool is eight cross-vendor, eight same-vendor and
four \texttt{astra} readings, and the same-vendor route gains 1.8 points from its first reading to
its eighth against the cross-vendor route's 19.3. It is sent temperature $0$ and returns nearly the
same verdict each time, though this design cannot say how much of that is the temperature and how
much the model. Counting that route as roughly one reading puts the depth nearer thirteen than
twenty (a heuristic, not an estimate), and the pooled figure should be read as a union over three
routes rather than over twenty independent looks.

The price is measured on the same footing. Union false positives on 150 correct increments reach
16.0\% [10.1, 22.3] at eight readings. The recall bought per point of false positive, meaning
recall gained over one reading divided by false positives gained over one reading, is
1.68 at eight, and it changes little as readings accumulate: from $K=2$ to $K=8$ the
ratio runs 1.67, 1.63, 1.64, 1.65, 1.66, 1.67, 1.68. It falls once, at $K=3$, and then rises, so repeated
reading on this route buys recall at a slightly improving price. The ratio is a quotient of two
measured rates and carries no interval of its own.

\subsection{Which model reads, and under which severity rule}
\label{sec:auditor}

Replacing the auditor with a stronger model of the generator's own vendor moved union recall at
eight readings by $-26.4$ points, cluster $[-37.3, -15.6]$: 4 of 110 against 33 of 110. Three
instances were flagged only by the new family and thirty-two only by the shipped auditor, at an
exact McNemar $p$ of $4.2 \times 10^{-7}$ and a cluster sign-flip $p$ of $2.0 \times 10^{-5}$.
This was the auditor-family study's single preregistered primary, so no family correction applies. Three
measurements qualify what it means, and none of them favours reading it as a statement about
auditing ability.

First, the two auditors are not at the same operating point, and this paper elsewhere warns against
comparing auditors at different operating points. The stronger model flags correct code at 3.3\%, or 5 of 150, which
\S\ref{sec:stats} says to quote as a count; its bootstrap interval, [0.7, 6.7], sits in the
range where simulations of an analogous interval cover about $0.89$ at a true rate of $0.03$
rather than $0.95$, so it may be narrower than it claims. The shipped auditor flags 16.0\% [10.1,
22.3]. The stronger model blocks less of everything.

\begin{figure}[t]
\centering
\includegraphics[width=0.46\textwidth]{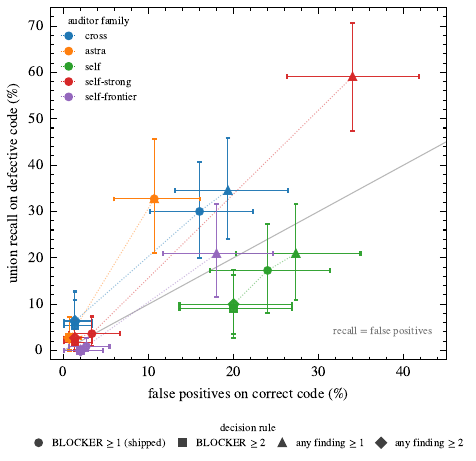}
\caption{\textbf{Union recall is a property of the decision rule, not only of the auditor.}
Each point is one auditor family re-graded under one of four decision rules on the same archived
readings, with problem-cluster 95\% intervals on both axes; no new model calls were made. The
families are read at different depths: \texttt{cross}, \texttt{self} and \texttt{self-strong} at
all eight of their readings, \texttt{astra} and \texttt{self-frontier} at their four. The figure
therefore places each family at its full archived depth and is not an equal-depth comparison; the
common-depth contrasts are in Table~\ref{tab:primary}. The dotted guides join a family's four
points and are not a measured curve. Two things Table~\ref{tab:primary} cannot show: the families
do not lie on one frontier, and \texttt{self-strong} has no point under the four deterministic
rules between 3.3\% and 34.0\% false positives, exactly where a comparison matched to the
cross-vendor auditor would sit, so its ranking there is bracketed rather than measured. Nothing in
this figure assesses finding content; a flag is not a diagnosis.}
\label{fig:sweep}
\end{figure}

Second, the sign reverses under a weaker severity rule (Figure~\ref{fig:sweep}). Counting any
finding rather than only findings graded BLOCKER, an exploratory rule that was not preregistered,
the stronger model reaches 59.1\% [47.3, 70.6] on the defect population at 34.0\% [26.5, 41.7] on
correct code, against the shipped auditor's 34.5\% [23.6, 45.5] at 19.3\% [12.8, 26.3]. The ratio
of those two rates, a derived point ratio with no interval, is 1.74 and 1.79 respectively, so
nothing here shows one family buying more recall per false positive than the other. Under
any-finding grading the stronger model's findings cover more instances, and under the BLOCKER rule
fewer, so the $-26.4$ depends on the counting rule. That does not show why the families' findings
differ, or that any finding identifies the hidden failure.

Third, the two families are not sampled alike. The shipped cross-vendor route's capability card
does not permit a temperature, so no sampling parameter is sent and the route varies between
readings; the same-vendor route permits one, the layer sends temperature $0$, and that route is
nearly deterministic. A union over nearly identical readings cannot grow much, and it does not:
over $K = 1$ to $8$ the cross-vendor family gains 19.3 points and the stronger same-vendor family gains
3.0. At one reading the gap between them is $-10.1$ points, not $-26.4$. The rest accumulates as
the two unions grow at different rates, and because model and sampling differ together, the
design cannot say how much of that difference is the sampling and how much the model. We regard
this as the most serious limitation of the contrast, and \S\ref{sec:limits} records what would be
needed to remove it.

A third family points the other way and belongs beside the headline. A high-reasoning model of the
cross-vendor family reaches 32.7\% [20.7, 45.0] union recall at four readings while flagging
10.7\% [5.9, 16.1] of correct code, against the shipped auditor's 30.0\% at 16.0\% with eight.
Paired on the same instances, the two advantages are never established together. At a common four
readings it leads on recall, $+11.5$ [2.9, 21.3], with an unresolved false-positive difference
of $-0.3$ $[-3.8, +3.4]$; against the shipped auditor's full eight it leads on false positives,
$-5.3$ $[-10.3, -0.7]$ unadjusted, but not on recall, $+2.7$ $[-6.5, +13.0]$
(\S\ref{sec:inreview}). The evidence therefore does not support ``a stronger model is a worse
auditor''. It supports the narrower statement that this particular same-vendor model, at this
the preregistered BLOCKER rule and sampled at temperature $0$, blocks far less.

A same-vendor route of the generator's own size reached lower union recall than the
cross-vendor one on these instances, $-12.7$ points $[-25.0, -0.9]$ at eight readings, signed so
that a negative value favours the cross-vendor route. The interval is unadjusted, its upper end is
0.9 points from zero, and that route is sent temperature $0$, so this is a direction under this
sampling and not a ranking of auditors. A study on a second substrate was meant to test whether the
sign reverses there; it closed without a quotable number, and on its clean run the same contrast
included zero, so no reversal was established either (\S\ref{sec:inreview}). The figure above is
a measurement on this substrate, and whether it is a property of same-vendor auditing is open.

\subsection{One rulebook sentence moves flags, and does not measurably move outcomes}
\label{sec:rulebook}

Adding a single rule to the auditor's constitution, telling it to find what the visible tests do
not cover, moved flags on the defect population from 10 of 56 to 25 of 56: $+26.8$ points, cluster
[13.6, 40.4], exact McNemar $p = 0.0003$, cluster sign-flip $p = 0.0009$. Together with the pooled
defect-and-correct flag contrast, this is one of only two comparisons that clear the Bonferroni
threshold, and the two are the same effect seen twice.

The estimand is narrower than that of the sections above and must be read at its own scale. It is measured
on 56 instances from 41 problems, not the 110-instance defect population. It is recall at a single
reading, union recall at $K = 1$ on that population, because every loop arm revises once. And it
is exploratory, because the preregistration named outcome contrasts rather than flag contrasts. For
those reasons we do not compare its magnitude with the $-26.4$ or the $+19.3$ of the preceding
sections, which are union-recall differences at $K=8$ over 110 instances (a contrast between
routes and a gain from $K=1$).

Two costs travel with it. On correct code the same rule moved flags by $+12.50$ points. Every
discordant pair points the same way (7 against 0), so by the rule stated in \S\ref{sec:stats} the
percentile bootstrap's bound at zero is an artefact, and the admissible interval is the
unconditional one. That interval, computed with a grid-approximated nuisance supremum and ignoring
clustering, is approximate rather than guaranteed exact; it is $[-3.6, +26.8]$ and
does not exclude zero, and its exact McNemar and cluster sign-flip $p$ are both $0.0156$,
against that study's Bonferroni threshold of $0.00313$. The directed statement the records
support is the pooled one, where 23 discordant pairs point one way and 1 the other (exact McNemar
$p = 3.0\times10^{-6}$, cluster sign-flip $p = 1.0\times10^{-5}$): the sentence makes the auditor
louder overall. On the correct stratum alone the cost is $+12.50$ points with an interval that
contains zero.

The second cost is to outcomes. On the contrast that isolates the rule, the arm with the added rule against
the cross-vendor arm on hidden-suite pass rate after one revision (both strata, 112 instances on
96 problems), the effect is $+5.36$ points, cluster $[-0.89, +12.07]$, exact McNemar $p = 0.146$
and cluster sign-flip $p = 0.183$, which does not exclude zero. The rule demonstrably moves what
the auditor writes down. Whether that converts into code that passes the hidden suite is not
established, and since this paper argues that a flag is not a catch, that distinction applies to
our own intervention first.

\subsection{Most of what the auditor missed is failure the specification never determined}
\label{sec:residual}

Fifty-seven of the 110 defective instances were flagged by no reading of the three families read
at that point. That is a statement about twenty readings, not about detectability. If an instance
is caught with probability $p$ on each independent reading, twenty readings miss it with
probability $(1-p)^{20}$, which is still $98\%$ at $p = 0.001$ and $36\%$ at $p = 0.05$. Never
flagged here does not mean never flaggable, and we use the residual as a population to
characterise rather than as a set of defects shown to be invisible.

We can put a number on how flaggable it is. Reading six families at forty readings, including the
same auditor under a rulebook with one sentence added, shrinks the never-flagged set from 57 to
32 of 110, and 43 of the 68 instances both raters call specification-undetermined
are flagged by some reading of some family. Being undetermined is not sufficient for being
missed. What does not move is the composition: the undetermined share of the never-flagged set is
44 of 57, $77.2\%$, at three families and twenty readings, and 25 of 32,
$78.1\%$, at six families and forty readings.

The first classification of that residual asked whether the visible tests exercised the failing
input before it asked anything else, and it called 46 of 57 unexercised edges. Re-rating the same
instances under a second protocol reverses that. The second protocol asks first whether the
specification's prose determines the hidden suite's expected value, redefines the categories by
entailment, and uses two raters in consensus; ordering, definitions and raters all changed
together, and the study cannot isolate which mattered. The two raters agree that 44 of 57
(77.2\%, cluster [62.1, 91.1]) are cases where the prose does not settle the value, and that 2 of
57 are cases where it does and the visible tests merely never built the input. The preregistered
condition for withdrawing ``the residual is mostly unexercised edge'' fires on both of its
clauses, and that claim is withdrawn. The interval is conditional on a category and has no
committed coverage simulation, so it is uncalibrated, as is every category-conditioned interval we
report.

A classification that reverses when its protocol changes is, by itself, weak evidence about the
auditor, and the raters are not independent of it: one was the agent that ran this programme
(\S\ref{sec:methods-review}), which wrote the rubric and knew the earlier labels and which claim
was at stake. Three things support the second reading. The rubric was fixed before any label was
written. The labels were committed before we looked for outside evidence. And when a first pass
was found to have carried instance identifiers and a sheet name into the raters' view, the
labelling was re-run with identities stripped and the result held. Where an outside party had
independently recorded that one of these specifications is defective, in twelve MBPP tasks named
in a paper written for another purpose \citep{richter2026collapse}, our raters agree on 9 of the 11 that fall inside our
defect population, with both disagreements on the boundary that paper's own ``incomplete''
category draws. That check is one-sided by construction: it can support agreement where an
outsider looked, and can bound nothing about the rate.

\textbf{A rater outside the residual's families, asked a different question.} One of the two
raters above is \texttt{gpt-6-astra}, whose own misses help define the residual. A model that is
not one of the families defining it, \texttt{gpt-5.6-luna} (which read these instances only as a
secondary audit route in an earlier architecture study), later rated the same 57 instances post
hoc under a different rubric: whether the prose determines an answer at all, on a sheet showing at
most three failing inputs and never the expected values (\S\ref{sec:inreview}). It labels
38 of 57 undetermined, 13 determined and 6 cannot-tell. The two figures measure different
constructs. The earlier category also admits prose that determines an answer the oracle
contradicts, which this rubric does not count as undetermined, so 38 is not a corrected estimate
of 44. What they share is that a participating and a non-participating rater both put a majority
of the residual on the undetermined side, and the second agreed with itself on only 7 of 11
repeated items.

\begin{figure*}[t]
\centering
\includegraphics[width=0.94\textwidth]{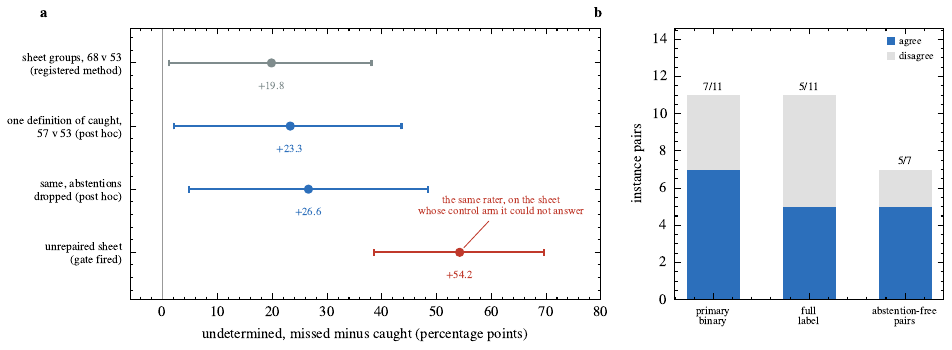}
\caption{\textbf{The residual contrast, printed beside the instrument that produced it.}
a, The share of instances a rater judged specification-undetermined, missed group minus
caught group, under each reading, with problem-cluster 95\% intervals. The two sheet groups use
different definitions of \emph{caught} and share eleven instances, so the two readings that apply
one definition are the ones that compare populations; both are post hoc. The red point is the same
rater on the same sheet before its control arm was rebuilt, when 42 of 53 caught entries carried
no failing-input evidence and all 42 drew \texttt{cannot-tell}. With that arm the contrast was
larger and pointed the way the hypothesis points. The repair and a fresh model run changed
together, so presentation, sampling and batch context are not separated, and a preregistered
one-third \texttt{cannot-tell} gate excluded the earlier figure. b, The same rater's agreement with itself on
the eleven instances the sheet carries twice, with identical text and identical evidence. The
contrast in a is read off the instrument in b, and the claim this paper draws
from it is stated under that constraint.}
\label{fig:residual}
\end{figure*}

\subsection{Nine further studies, and what they license}
\label{sec:inreview}

Eight further studies cleared independent review after between three and eleven rounds each, and a
ninth, on a second substrate, closed without a quotable number. We
report them here rather than folding them into the sections above because several are narrow or
negative results whose limits are part of the claim, and because the shape of their refusals is
itself a finding about this method.

\textbf{A rulebook sentence moves a second vendor's family.} The referent rule, \emph{find what the
visible tests do not cover}, raises union BLOCKER recall on \texttt{self-strong} (\texttt{claude-sonnet-4-6})
at four readings by +20.9 points, cluster [10.8, 31.8], at a false-positive cost on
correct code of +8.0 points [3.4, 12.7]. A second intervention, a rule on how findings are graded, shows no demonstrated
improvement, $+0.9$ points $[-1.8, +4.5]$, and that hypothesis fails its registered criterion,
which demonstrates neither equivalence nor that the rule did nothing. This reproduces
\S\ref{sec:rulebook}'s effect on a different vendor's model at a different $K$, with its own
price. A post hoc quantity from the same study, whether a flagged finding \emph{reports a failure}
on the instance's failing input class (says that class raises or returns the wrong value), is 3 of 110 under the shipped constitution and 27 of 110
under the rule, and it may not be quoted as defect recognition. Its rubric admits a finding even
where it grades the class non-blocking or says the specification is silent, so findings that
disclaim the requirement are counted in.

\textbf{The added rule raises union recall to 60.9\%; a higher ceiling is not established.}
On the defect population the same rule takes union recall (the identical quantity: blocked by at
least one BLOCKER, unioned over eight readings) from 33 of 110 to 67 of 110 = 60.9\%
[48.6, 72.5], +30.9 points [19.1, 43.1], at +20.7 points [12.8, 28.8] of false
positives, with a single-draw false-positive rate of 23.6\% [17.1, 30.3] that fails the product bar on all eight
draws. The caution that a flag is not a catch attaches to both figures equally. What separates
them is their price, 36.7\% [28.5, 44.8] false positives against 16.0\%, which is why the shipped
route is the shipped route. The fitted asymptotes differ by $+27.1$ points [10.9, 39.0], and
\emph{that does not establish a higher ceiling}. The cross-vendor arm never flattened, so one of
the two terms is an extrapolation. Heterogeneous low-probability detection imitates a ceiling at
this budget, since an instance found with probability $0.02$ per reading survives eight readings
about 85\% of the time. And this study's own zero-inflated estimator finds no never-detectable
class in \emph{either} arm. Its adjudication kill is arithmetic rather than a labelling artefact:
only 9 of 110 instances drew any finding on the single reading whose finding texts were archived, so the largest defect-asserting count available under
perfect adjudication was 9, below the floor of 20. Every adjudication-derived figure in it is
conditional on labels whose production can no longer be audited, because one rater's raw reply,
prompt, launcher and execution log lived only in a session scratchpad and were destroyed between
the review that flagged them and the repair.

\textbf{Corroboration does not validate a generated test.} A wrong generated test, one that fails
on the dataset's canonical solution, is not reliably recognised by asking whether further
independent generation draws also fail on the same candidate. Neither rule that makes further generation calls reaches the
preregistered threshold on the registered population, which covers failing applications on all
three candidate strata, 58 of those 107 failing applications on candidates that already fail their visible
tests. The retained wrong-test rates are 12.2\% [3.6, 22.6] (Wilson [7.0, 20.6]) for majority-over-draws
and 14.9\% [6.5, 25.5] for any-draw agreement, against a limit of 2\%, and neither is shown to improve on the free
within-draw comparator at 12.8\% [3.9, 23.9]: the better of the two rules keeps the same wrong applications plus four
correct ones, a small observed advantage the study cannot resolve. On the candidates a product
would meet, those passing their visible tests (post hoc), the rates are 6 of 33, 18.2\%
[0.0, 44.8], and 10 of 43, 23.3\% [6.8, 47.1]. The first leaves its relation to the 2\%
threshold unresolved; the second's unadjusted interval lies wholly above it, which is exploratory
evidence of exceeding it. Neither is a preregistered test on that population, whose registered
kill rule was a point estimate on the mixed population. A floor of 2.4\% holds on the original
unique-test quantity because 29 of the 45 wrong tests never fail on any candidate at all. The
claim is negative and bound to this substrate: these rules do not validate generated tests here,
which is not to say generated tests cannot be validated.

\textbf{An injected population does not give the residual result a prospective test.} We
constructed a population of injected defects intended to make specification-determinedness hold
by construction. It does not. The filter registered to recover each edit's first failing hidden
input is a truthiness check on a field the injector wrote: of 281 recorded witnesses, 179 are
strings and 102 are lists, and none is executed. Four of the remaining five filters accept cases
their registered wording excludes, such as a suite that reports success after an early exit, a
process that terminates without an assertion or an exception, and an added import from a module
already imported. No filter establishes specification entailment, and repairing the first one
alone would not have: recovering the failing input and executing a witness shows that the edit
changes behaviour on an input the hidden suite exercises, not that the specification entails the
value expected there. What may be quoted is descriptive. At eight readings the auditor blocked
90 of 92 accepted injected instances against 11 of their 92 unmodified twins,
79 discordant pairs one way and none the other, and against 33 of 110 on the natural
residual. The instances the construction's own gate \emph{refused} were blocked 31 of 40,
which is why that gate cannot be assumed conservative. A probe separates the injected set from the
natural one at 96.7\% on the items it answered, \emph{by a feature it does not identify}: the two
arms share no problem, so task composition alone could produce the result, and the probe's model
is one of the two gate models. \S\ref{sec:residual} therefore keeps exactly the status its own
study gave it, post hoc and unreplicated. A successor needs an independent party to validate the
actual violation \emph{and} the specification-entailing expected behaviour under a fixed, blinded
protocol. Neither a second substrate nor an independent rater suffices alone, because task
composition and construction confounding survive both.

\textbf{A study on a second substrate closed without a quotable number.} In a second benchmark,
the visible-test text shown to the auditor \emph{and} to the generator was sliced out of its test
class without the class header, so it failed to parse on all 300 tasks while scoring executed the
intact class. The repair for that fault introduced a second one, re-attaching a hidden method's
decorators to the first visible method on 26 tasks. A third run on a clean extraction reproduced
five of the six reversals of first-substrate findings that the second run had produced, and did not
reproduce the sixth. The
criterion registered before those numbers existed was stated over the set, so the strict reading
applies and the study is closed after three runs, about \$82, and no quotable number. Whether five
of six should have licensed the five is recorded as an open question in \texttt{CLAIMS.md}.

\textbf{Measured union recall is rule-dependent, and a headline's magnitude belongs to its own
rule.} Re-grading findings already archived by two of the studies above under four decision
rules, with no new model calls, leaves the shipped BLOCKER threshold's estimate intact for its own
rule, which \emph{was} preregistered as ``$\geq 1$ BLOCKER finding'', while \emph{reversing its
sign} under any-finding grading. That reversal is a mechanical consequence of changing what
counts, and it identifies nothing about why the families produced different findings or
different severities. At a common depth of four readings, \texttt{astra} exceeds the
cross-vendor auditor on union recall over the defect stratum by +11.5 points [2.9, 21.3], paired within instance, with a false-positive difference of $-0.3$ points
$[-3.8, +3.4]$. The two sides are not averaged alike. The cross-vendor arm has eight readings and
is averaged over all $\binom{8}{4}=70$ four-draw subsets, while \texttt{astra} has exactly four,
so $\binom{4}{4}=1$ and its estimate is a single realisation with none of its reading-selection
variance averaged out; one arm's luck in four draws is fully in the number and the other's is not.
On the cross-vendor auditor's complete eight-draw ladder the same contrast is $+2.7$ points
$[-6.5, +13.0]$ with a false-positive difference of $-5.3$ points $[-10.3, -0.7]$. That
comparison points toward lower false positives, not higher recall, on an interval whose upper end
is 0.7 points from zero and which is not adjusted for the other contrasts computed beside it, so
it is a direction the data lean and not one they establish. Observed false-positive rates near one
another are not a matched operating point, and one family has no point under the four
deterministic rules between 3.3\% and 34.0\% false positives, precisely where a matched comparison
would sit, so its ranking is bracketed rather than measured. Nothing here assesses finding content
against the hidden failure, so none of it speaks to what any finding identified.

\textbf{The ambiguity gap survives a rater that did not write the rubric, and the instrument that
shows it disagrees with itself.} The contrast in \S\ref{sec:residual} between missed and caught
instances had been rated by two models, one of them the agent that wrote the rubric. A third model
re-rated the whole sheet after we rebuilt it on mechanically recovered witnesses. That rebuild was
necessary because the control arm had been unanswerable: 42 of 53 caught entries carried no
failing-input evidence, and all 42 drew \texttt{cannot-tell}. The share labelled
\emph{undetermined} is 43 of 68 against 23 of 53, +19.8 points [1.2, 38.1]. Those two
groups are not two populations, because they define \emph{caught} differently and 11 instances
fall in both; removing those 11 from the missed group gives $+23.3$ points [2.0, 43.6], post
hoc and larger, so the overlap had understated the contrast. The limit is part of the result. On
those 11 instances, shown with identical text and identical evidence, the rater agreed
with itself on 7 of 11 on this binary outcome and 5 of 11 on the full label, where the earlier
study's two raters each agreed 11 of 11 (Figure~\ref{fig:residual}). Every interval's lower end
lies within five points of zero, and this remains a model; no human has rated this material. The
\emph{unrepaired} sheet scored $+54.2$ points [38.5, 69.5], 2.7 times the repaired contrast
and in the direction that flatters the hypothesis. That is a descriptive association between one
repair and one fresh re-rating, not evidence about specifications: the repair, the run and its
sampling changed together, and the preregistered one-third \texttt{cannot-tell} gate catches it.

\textbf{Adding the missing rule to the specification moves diagnosis, but only where the rule
matched the oracle.} The contrast above conditions on the specification; this study changes it.
The population is the 44 never-flagged instances that both earlier raters called
specification-undetermined. For the 32 on 19 problems that passed the generation gates, a model
shown the failing \emph{inputs}, but never their expected values, wrote one behavioural rule into
the prose, and an auditor re-read the same code and the same visible tests under the original, the
clarified and a same-length placebo specification, four readings each. An instance counts as
diagnosed only if both adjudicators judged some finding to state the behaviour the hidden suite
expects. Diagnosis rose from 0 of 32 to 9 of 32 under clarification and stayed at 0 of 32
under placebo. The registered primary, clarified against original, is +28.1 points,
problem-cluster 95\% interval [7.7, 51.4], which meets the registered criterion. The registered secondary, on the 23 instances where the re-rating study's consensus (\S\ref{sec:residual}) and
the third rater's label agree, is 8 of 23 against 0 of 23, $+34.8$ [10.0, 60.9].

Given the observed data, that criterion added little. Every discordant pair points one way, so no
bootstrap resample can fall below zero, and with successes on five of 19 problems a resample is
exactly zero only when it misses all five, with probability $(14/19)^{19} \approx 0.003$; the
registered run has 27 such resamples in 10{,}000. Once five success-bearing problems are observed,
the interval's exclusion of zero adds little to them (with one such problem it would not have
excluded zero), and \S\ref{sec:stats} says to read a one-signed contrast from an exact test
instead. The registration kept the bootstrap and we report it as registered, but the informative
figure is the cluster sign-flip test the registration prefers. The nine successes sit on five
problems, and the sign-flip test gives $p = 0.060$ (exactly $2/2^5 = 0.0625$ on enumeration), against $0.0039$ for
a McNemar test that counts nine independent pieces of evidence where five exist. It is not
significant at 0.05. The registered naming-rate secondary was never computed.

Two further limits are part of the result. \emph{The inferred rule contradicts the hidden suite
entirely on 23 of the 32 instances and partly on 2}, and diagnosis was 0 of 23 where it
contradicted against 9 of 9 elsewhere. That split was classified after the outcome was seen, not
assigned, and the agreeing instances may simply be the easier problems; it does not say which
property of the edits moved diagnosis. And a separate model rated the clarified prose more
determining than the original by $+28.1$ points [2.6, 54.3] but than the placebo by $+18.8$
$[-9.4, +45.2]$, so whether the placebo was inert as a specification is unresolved. One
adjudicator is the agent that ran the study and the other is \texttt{gpt-6-astra}, the model that
also reviewed this study (\S\ref{sec:methods-review}); they agreed on 84 of 91 items,
$\kappa = 0.85$, which is agreement and not accuracy. The study supports a specification-edit
contrast on these selected instances. It does not establish that underdetermination caused the
original misses, and \S\ref{sec:residual} is not strengthened into that claim by it.

\textbf{What a flag on correct code points at.} The false-positive rates above count flags on code
that passes its hidden suite. A suite may leave behaviour unexercised, so a flag there need not be
a false alarm. A preregistered study re-read the 150 correct instances with the shipped route at
eight readings, keeping the finding texts the ceiling study had discarded; 20 were flagged ($13.3\%$
[7.9, 19.2], a fresh sample of the route's $16.0\%$). A model given each finding, but not the
reference solution or the hidden tests, extracted the concrete inputs it names, and candidate and
reference were run on them. Two of the 20 disagreed with the reference on an extracted
input the hidden suite does not exercise ($10.0\%$, problem-cluster [0.0, 26.3]); on nine they
agreed, and nine yielded no valid input. Because the interval includes zero, the registered
reading is that at this resolution there is no evidence that this route's flags on test-passing
code identify reference disagreements. No class licenses the words ``true positive'' or ``false
alarm'': a disagreement does not show that the specification was violated, and agreement on an
extracted input does not show a finding wrong.

\textbf{What the refusals were about.} These eight studies cleared review at rounds 3, 3, 4, 6, 7,
9, 10 and 11, fifty-three rounds in all. Most of those rounds ended in a refusal of the
\emph{repair} rather than of the measurement. Measurements moved on review in three of the eight:
the sweep's first review found its loader reading one family's draws 4 to 8 as 1 to 5 and
another's population from a partial draw, and every cross-family figure was recomputed; a label
count in the third rating's report was 71 where its data say 67; and the clarification study's
primary moved by one instance and three classification rows (\S\ref{sec:methods-review}). Every
figure above reproduced independently at every round after its last repair. The recurring failure
was a check or a sentence that claimed more than it delivered. A binding protected one sentence of
a withdrawal spread across six; when it was rewritten to protect all six, two of them were
truncated before the words that carried their meaning. One check stripped HTML comments and called
the result what a reader sees, and one test was named for a code path that never called it. A
second pattern appeared in the last thirteen rounds: most of those refusals were
\emph{overstatements away from our own interest}, such as a registration whose chronology its own
repository contradicted, a withdrawal that overcorrected past what the record supported, a
publication prohibition a registered gate never imposed, and a limitation credited to another
study's work that the other study had not performed. Understatement was corrected as
often as overstatement; in both cases the fix was a smaller claim written where the claim was.

\subsection{What follows for a recall number}

Taken together, the results above say an audit recall figure is a joint property of the auditor
and its material. It moves with which route reads (model and sampling together, which this design
does not separate), with what the rulebook tells it to look for, and with the severity threshold
at which a finding becomes a block, which is rarely reported. What it misses is associated with
whether the hidden suite asks a question the specification answers. That association is observed
in the residual and in a specification-edit contrast on selected instances; it is not isolated as a
cause.

\section{The science studies in full}
\label{sec:ai4s-full}

The general-code studies (Section~\ref{sec:ceiling}) measured the model tier on short general-purpose functions. A scientist would hand
an auditor three different jobs: check the code that computes a result, check that a reported
result follows from that code, and check the data an analysis starts from. We ran one
preregistered study per job, each with ground truth that no model supplies. Scientific code
(\S\ref{sec:ai4s-code}) is judged by scientists' numeric tests, results (\S\ref{sec:ai4s-results})
by re-execution, and data (\S\ref{sec:ai4s-data}) by the faults we injected. Every item in all
three is audited through the product's own audit call with its shipped rulebook, and the same
flag rule holds throughout: an item is flagged when the model grades any finding BLOCKER.

\subsection{Scientific code}
\label{sec:ai4s-code-full}

\textbf{Substrate.} SciCode \citep{tian2024scicode} decomposes 80 research problems from 16
subfields into 338 steps, each a function with numeric tests written by scientists. Its targets
are released separately from the code, which lets the tests act as a hidden oracle. On the
benchmark's development split, the gold code passes 48 of 50 steps in our environment; the two
steps whose gold fails here are excluded, as are the three the official harness skips.

\textbf{Population.} The generator of the general-code studies wrote every step in SciCode's own protocol,
three samples per step, each sample building on its own earlier steps. A candidate is
\emph{correct} if it passes the step's tests and \emph{defective} if it fails or times out. A step
enters either stratum only if at least one sample passes it, so a failure counts only where this
environment has shown the step can be passed. This yields 81 defective instances from 30 problems
and 374 correct ones, from which 150 (53 problems) were drawn by seed.

\textbf{What differs from the general-code studies.} SciCode has no visible and hidden split, and we did not invent
one. The auditor sees the problem, the step, its background, the function header, the allowed
dependencies and the candidate, and no tests of any kind. The shipped cross-vendor auditor and the
generator's own model each read every instance eight times. The generator's model is reached
through a command-line route rather than the API; the route adds a short fixed context and uses
default sampling, so this study's same-vendor family is not temperature-0, unlike theirs.

\textbf{Two arms.} After the registered arm's outcomes, a second registered arm changed only the
task text, naming the step's function as the deliverable; the program bytes are identical.

\textbf{Residual.} Defective instances flagged by no reading of either family are rated on a
blind sheet under the general-code residual's question (\S\ref{sec:residual}) adapted to science: do the step's prose and background
determine the value the test expects, including tolerance, units, conventions and numerical
method. The raters are the agent that ran the study and a model that audits nothing here.

\textbf{Results.} In these runs, how the task framed the deliverable changed the flag rate
substantially (Figure~\ref{fig:ai4s}a). With the whole problem as the task, the shipped auditor
flagged 87.7\% [72.7, 98.6] of defective and 68.7\% [56.6, 79.7] of correct instances at eight
readings; some of its objections to correct code were that one step did not implement the whole
problem. With the task reframed around the step, flags on correct code fell to 36.0\% [26.4, 45.6], a
paired change of $-32.7$ points [$-42.8$, $-22.4$], and recall fell to 72.8\% [59.3, 87.3], $-14.8$
points [$-28.6$, $-2.7$], though the cluster sign-flip test for that recall change gives
$p = 0.0625$. The change is the whole reframing (the step named, the problem description moved last
and marked as context, earlier functions exempted), and the arms ran in sequence. The generator's own
model's recall was 46.9 points lower than the shipped auditor's in each arm ([$-63.0$, $-30.7$] and
[$-61.5$, $-34.4$]; a route comparison), and it flagged 27.3\% [18.4, 37.3] and 18.0\% [11.5, 25.6] of
correct code, respectively.
At one reading in the reframed arm the shipped auditor flags 61.7\% [47.9, 75.9] of defective against
19.8\% [12.9, 27.2] of correct instances: it discriminates, modestly. A read sample of 20 of its flags
on correct code ranges from real divergences between code and step text to contradictions inside the
benchmark's own instructions and tests too weak to separate outputs; which flags point at defects was
not established. An independent expert audit of SciCode's test problems found self-contradictory
specifications, non-reproducible gold answers and over-tight tolerances that reject correct solutions
\citep{hu2026scicodeverified}, so some instances labelled defective here may be correct code the
benchmark rejects. Of the 81 defective instances, 7 and 19 escaped all sixteen readings of
the two families in the two arms; both raters judged most of them (4 of 7, 13 of 19) undetermined by
the prose, but the registered test that this share exceeds the
general-code residual's 77.2\% is killed in both arms by its own interval rule.

\subsection{Scientific results}
\label{sec:ai4s-results-full}

\textbf{Items.} Each correct SciCode instance whose output on one seeded test case is a real
scalar or a real array of at most ten elements becomes a small results increment: the program, the
test's inputs, a script that runs one on the other and writes a run log, a \texttt{results.json}
citing each value's log line, metadata declaring every input at a revision, and a one-paragraph
report whose numbers carry the product's number$\to$source annotations. Seventy-four of the 150
correct instances qualify, from 36 problems. Each yields one clean item and one faulty item. The
faulty item carries one of seven faults, assigned in rotation by seed. Five are \emph{reporting}
faults, where the artefacts disagree with each other or with the log the code wrote: a $\times
10^{3}$ scale slip, a sign flip, a value from another test case, a 5\% relative error, and a
transposed digit pair in the report alone. Two are \emph{fabrications}, where the scale slip or the
5\% error is written consistently into the log, the results and the report, so that every artefact
agrees with every other and none agrees with the code.

\textbf{Three auditors.} The LLM auditor reads the increment four times, with the project's checks
set to the shipped \texttt{science} profile. The model's prompt includes the deterministic
result, as it does in the product, so on reporting faults the model has seen the deterministic
finding. The deterministic profile's own verdict comes from the same call. Re-execution runs the
program on the inputs and compares the output with the reported values.

\textbf{Gate.} Before any model call, every clean item had to pass the science profile with no
BLOCKER and re-execute to its reported value; all 74 did when first built. A two-reading pilot then
found an instrument defect: no file in the item performed the run the report described, and the
auditor blocked on that, correctly. We added the run script, discarded the pilot readings, and
added a third condition to the gate, that each clean item's run script reproduce its run log byte
for byte. All 74 passed all three conditions with no repair.

\textbf{Results.} The deterministic profile does exactly what it was built for (Figure~\ref{fig:ai4s}b).
It flags all 49 reporting faults, because the log, the results and the report no longer agree and
number$\to$source says where, and none of the 25 fabrications, because in those everything agrees
with everything else. It flags no clean item. The LLM auditor flags 23 of the 25 fabrications at
four readings, 21 of them with a sound finding about the fault (post hoc). Those findings work the
result out from the code and inputs, for example evaluating a KL divergence of two small
distributions to 1 where the report says 1000. The other two rest on calculations that are
themselves wrong. Paired on the faulty items, all 23 discordant items go the LLM's way (exact
McNemar $p = 2.4\times10^{-7}$; a problem-level sign-flip test, post hoc, gives $3.8\times10^{-6}$).
On the reporting faults its rate is of the combined workflow, since its prompt carried the
profile's findings. Re-execution flags every fault except four whose change to a near-zero output falls within
\texttt{np.allclose}'s default absolute tolerance; two of those outputs are floating-point zeros.

The LLM also flags 12 of the 74 clean items, and on reading each (post hoc) 8 of those flags are right. In 9 instances the
benchmark passes the test's inputs inside the call itself, so the report's sentence that the
program ran ``on the inputs in \texttt{inputs.py}'' was false; 7 of those clean items were flagged
for it. Two of four outputs whose unit the solution documents, one of them also among the seven, were flagged for our
\texttt{dimensionless} label. The remaining 4 flags rest on wrong calculations or, in one case, on
a syntax error that is not there. On this evidence a provenance chain and a reader that recomputes
are complements: the first is exact about agreement and blind to consistent error, and the second
sees consistent error but is sometimes wrong itself.

\subsection{Scientific data}
\label{sec:ai4s-data-full}

\textbf{Substrate.} We use four physical-science tables from the UCI repository, each under CC BY 4.0:
concrete compressive strength \citep{yeh1998concrete}, airfoil self-noise
\citep{brooks1989airfoil}, combined-cycle power-plant output \citep{tufekci2014ccpp}, and
superconductor critical temperature \citep{hamidieh2018superconductor}. Exact duplicate rows and
physically impossible rows (relative humidity above 100\%) were removed from each pool before any
item was drawn.

\textbf{Items.} Each item is 120 rows sampled from one pool with a data card that gives every
column's meaning and unit and, where the source states one, its range. There are 35 clean items
per dataset and five per dataset for each of seven faults: mixed units in 20\% of one column's
rows, three negated values of a non-negative quantity, 12 duplicated rows, the target shuffled
among 30\% of rows, two columns with disjoint ranges swapped, a $-999$ sentinel in 5\% of one
column's cells, and a continuous column rounded to integers.

\textbf{Auditors.} Both model families read each item four times, with the dataset as the
deliverable and the card as its documentation. The comparison is a deterministic validator written
from the card alone and frozen before any item was built: declared ranges, non-negativity,
duplicates, sentinels and integer-valued continuous columns. By construction it catches the
negations, duplicates, sentinels and rounding. It sees mixed units or swapped columns only when
they push a value below zero or out of a declared range, or make a continuous column
integer-valued, and a shuffled target not at all. The question is whether the model catches
what the validator cannot. Two instrument defects were found and fixed by amendment before the
readings reported here: the card and file named one column differently, and the files first sat
outside the audited scope, so the product had told the model there was nothing to audit.

\textbf{Results.} The two tiers flag overlapping but different faulty items (Figure~\ref{fig:ai4s}c). The validator flags
92 of 140 faulty items (65.7\% [57.9, 73.6]) and no clean item. The shipped auditor flags 65 of 140
(46.4\% [38.6, 54.3]) at 4 of 140 clean items (2.9\% [0.7, 5.7]), and their union flags 112 of 140
(80.0\% [73.6, 86.4]) at the same 4. (These intervals resample items within dataset; four datasets
are too few to resample, so no interval here carries between-dataset variation.) Of the 20 faulty
items the auditor flags and the validator misses, 18 carry a finding about the injected fault
itself, 9 of them mixed units and 9 swapped columns (post hoc). In one, the auditor reasoned that
three atoms spanning a 135\,pm radius range cannot have a mean radius of 1.6\,pm, which is the cell
the fault had converted to {\aa}ngstr\"oms. The other 2, and all 4 flags on clean items, are requests for
fuller column descriptions on one dataset's card; so are 2 more of its faulty-item flags, which the
validator also caught. Counting only findings about the injected fault (post hoc), the auditor flags
61 of 140 and the union 110, at no clean item. The auditor flagged no shuffled target and no
duplicated rows (0 of 20 each), and on every fault the validator was written to catch it did worse
than the validator. The generator's own model, read through its command-line route, is not a usable
data auditor here: it flags 67 of 140 clean items, and in at least 48 of those one finding states a
row count, field count, missing column or truncation that the file contradicts (post hoc lower
bound).

The deterministic validator is not neutral. It was frozen from the card before any item existed,
but it omits one range the concrete source's documentation states and uses a list of continuous
columns the card does not give; with the omitted range it would flag 93 of 140. Unit conversion
also left floating-point tails that mark some converted cells in two of the tables. None of the
auditor's mixed-unit findings cites them, but that does not show they had no influence.

\section{Discussion in full}
\label{sec:discussion-full}
\subsection{What a union-recall number means}

A union-recall figure depends on the false-positive rate it cost, so recall quoted without that
rate cannot be carried to another setting. We make this as a methodological point. The measured
support we tried to give it came from a second substrate that was run three times and closed
without a quotable number (\S\ref{sec:inreview}). The point needs no second substrate:
recall and false positives move together with any severity threshold, and a critic can be tuned
to any recall by blocking more. Our records support a narrower version on the substrate we
report. The same auditor at two severity rules sits at two different operating points
(\S\ref{sec:auditor}), and the shipped auditor's 30.0\% [20.0, 40.7] recall is quoted only beside
the 16.0\% [10.1, 22.3] it cost.

The cleaner experiment for the general claim is a severity sweep within one substrate. Our sweep
re-grades archived findings at four decision rules and shows each family's recall and
false-positive rate moving with the rule (Figure~\ref{fig:sweep}). Its rules are deterministic
and do not place the families at a common operating point. A randomised mixture of two archived
rules could reach one in expectation without new model calls, but it would be a new post hoc
policy needing its own uncertainty analysis, and we have not built it. Two substrates differ in
specification length, solution length, problem population, defect distribution and suite
construction all at once, so comparing them could never have isolated the operating point.

\subsection{Repeated reading buys nothing where the verdict does not vary}

Union-of-$K$ is the instrument this paper uses to look for a ceiling, and part of what it cannot
measure follows from arithmetic alone. A union over readings that do not differ cannot grow, so
the curve is flat for any route that returns the same verdict however often it is asked.

Our records establish the design fact that makes this matter here. The same-vendor routes are sent
temperature $0$, while the cross-vendor route is sent no sampling parameter
(\S\ref{sec:auditor}). From $K = 1$ to $8$ the first family's union gains 3.0 points and the
second's gains 19.3. No temperature-matched arm was run, so we cannot say how much of that gap is
sampling and how much is the operating point. We treat it as a confound on our own contrasts.

For best-of-$K$, union-of-$K$ or panel aggregation over critics, what follows is the arithmetic: a
route whose verdict does not vary between readings gains nothing by being read again. We measured
per-instance flag counts, not any distance to a decision boundary, and ran no panel aggregation,
so we offer no law for how much a varying route gains. The diminishing shape of the curves is not
itself evidence of a ceiling either. Exact subset averaging makes each reading's marginal gain no
larger than the last by construction, and the curves measure the size of those gains.

\subsection{What the ceiling is made of}

We do not rank these contributions, because they were measured on different populations, depths
and rules. Two of them share a scale. Eight readings of one auditor buy 19.3 points of union
recall. Twenty readings across three families buy a pooled 48.2\% [36.7, 60.0] at 36.0\% false
positives on correct code, and leave the complement, 51.8\% [40.0, 63.3] of the defect
population, untouched. Changing which model reads moves the same quantity (union recall at $K=8$
over the same 110 instances) by $-26.4$ points [$-37.3$, $-15.6$], a larger magnitude than repetition's 19.3, with the stronger
route flagging 3.3\% [0.7, 6.7] of correct code against 16.0\%. A same-size same-vendor
substitution moved it by only $-12.7$ [$-25.0$, $-0.9$], a direction under temperature-0 sampling, and \S\ref{sec:auditor} gives three reasons not to read either
number as auditing ability. Changing one sentence of the rulebook moves flags by 26.8 points
[13.6, 40.4], an exploratory contrast on 56 instances at one reading, at $+12.5$ points on correct
code ([$-3.6$, $+26.8$]) and with no established effect on hidden-suite outcomes; it is not on the
scale of the other two.

A fourth contribution is not the auditor's at all, and we can describe it only post hoc. Under a
rule that asks whether the specification's prose determines the hidden suite's expected value, 44
of the 57 never-flagged defects fail. We tried to build a population where the prose does
determine the answer. Review established that no filter in it enforces that property, so it
licenses description only (\S\ref{sec:inreview}).

The association between never being flagged and never being determined is empirical. Selecting
the missed instances does not fix their labels, and against caught instances the third rater's
post hoc contrast is $+23.3$ points [2.0, 43.6]. That rater agreed with itself on only 7 of
11 repeated items, and the interval is conditional on its recorded labels. Selection and
confounding rule out a causal reading, so the contrast locates the misses and does not estimate
how much the benchmark contributes.

One intervention changes the specification instead of conditioning on it
(\S\ref{sec:inreview}). Adding a behavioural rule to the prose moved diagnosis on 9 of 32 such
instances, all of them where the added rule did not contradict the hidden suite and none of the 23 where
it contradicted it. This is the only evidence here that manipulates the specification. Its
population was selected for having been missed and called undetermined, and it cannot separate a
specification that settles the question from one that settles it correctly.

The constructed population has two further limitations, one measured and one found in review. A
frontier model separates the constructed edits from natural ones at 96.7\% on the items it answered, and
between 79.9\% and 97.3\% once the items it would not classify are counted at either extreme. It
does so \emph{by a feature it does not identify}: the two arms share no problem, and the model is
one of the two that gated the population. And because the construction does not enforce
specification-determinedness, it licenses \emph{no} bound on the measured ceiling.

\subsection{Flagging is not naming}

An audit that blocks an increment for the wrong reason has still blocked it, and a recall number
counts that as a catch. For one reading of the shipped auditor the finding texts were kept and
adjudicated blind by two raters. The auditor returned a finding on 9 of 110 defective instances
and asserted the actual failing behaviour on 5 to 7 of them, depending on how the two raters are
combined. The raters agreed at $\kappa = 0.391$. We report that beside the rate because weak agreement
is part of what the rate is worth. That study's preregistered kill condition fired on
this result, and \S\ref{sec:inreview} states its headline in the form the kill requires. The
figure describes one reading. It is not a naming rate for the eight-reading union, which would
need the texts of all eight readings, and those were not archived.

\subsection{What the science studies add}

The science studies give the science profile's design rule (the model names
evidence, code verifies it) an empirical reading and a boundary. On results, number$\to$source
flagged every inconsistency between a report, its results file and its log, and could not see a
result that was wrong consistently; the model flagged 23 of 25 such fabrications, 21 with a sound
finding about the fault (post hoc). On data, a validator written from the card flagged what a rule can
state and missed what needs reading across columns and units; of the 28 such items the validator
missed, the model flagged 18 with a finding about the fault (post hoc).
Where both tiers ran, the deterministic one flagged no clean item, and every clean-item flag and every wrong finding
came from the model;
on data the union was better than either part, and on results the model's flags already contained
the profile's, whose findings its prompt carried. Neither study shows that the model tier scales: the
artefacts were small, the model's fabrication findings are hand calculations (two of the 23 rest on wrong ones), and it
missed every fault that needs comparison across many rows. A model reading is a useful second tier
where the artefact fits in the reader's working memory, not a substitute for executing the code or
computing over the data.

\textbf{Clean items must be clean by the auditor's standard.} In the results and data studies the
auditor flagged defects in our own construction: a card and file that named a column differently, a
file outside the audited scope, a report claiming a run no file performed, a provenance sentence false
for inputs passed as literals, and a unit label contradicting the code's documentation. Each was a
correct flag on an item we had built as clean. We caught two in pilots, one because every verdict
came back as an escalation, and the other two by reading every flag on a clean item, which the registrations
required. In the code study the task we wrote named the whole problem when one step was delivered,
and a sample of the auditor's flags on correct code objected to exactly that; there we read a sample,
not every flag. Without such reading, a false-positive rate can measure the evaluator.

\subsection{Limits}
\label{sec:limits}

\begin{enumerate}\itemsep1pt
\item \textbf{Sampling differs between families.} One route is sent temperature $0$ and the other
  no sampling parameter, which bounds how far one union can grow before any difference in ability
  is measured. Every cross-family union-of-$K$ contrast on general-purpose code inherits this,
  and it is the confound we would fix first. In the science studies the same-vendor route runs at
  its default sampling through a command-line route, a different asymmetry.
\item \textbf{No matched operating point.} No auditor is compared with another at a matched
  false-positive rate. Our sweep re-grades archived findings at four deterministic decision
  rules, under which one family has no point between 3.3\% and 34.0\% false positives. We built
  no randomised policy between them, so the paper cannot say whether the families differ at the
  same operating point.
\item \textbf{Interval coverage.} The primary general-code intervals are problem-cluster
  percentile bootstraps, with cluster counts that depend on the study and stratum (56 problems for the
  defect population, 143 for the correct sample); the science studies' have their own, and the data
  study resamples items within dataset. A simulation of an analogous single-rate interval (600 resamples
  rather than 10{,}000, tested rates only) covers near nominally at the headlines' rates and
  at 0.80 at the one tested rate below $0.10$ (0.03; \S\ref{sec:stats}). For the paired contrasts and
  conditional shares coverage is unmeasured, and no sample size was justified in advance by a
  detectable-effect calculation.
\item \textbf{Substrate.} The substrate is short, self-contained Python functions with executable
  hidden tests, so nothing here measures auditing a change to a large program with no oracle.
  Both benchmarks are old enough that memorisation cannot be ruled out.
\item \textbf{One generator.} The defect population is one model's failure mode.
\item \textbf{Raters.} The residual rubric's raters were two models, one of them the agent that
  wrote the rubric, which was not blind to the earlier labels. One external label set overlaps eleven of
  the rated tasks, and the raters' labels agree with it on nine.
\item \textbf{The constructed population.} It did not enforce specification-determinedness, and
  its edits are separable from natural code at 96.7\% of the items the probe answered
  (79.9--97.3\% once unanswered items are counted at either extreme) by something the probe does
  not identify.
\end{enumerate}
Each limit is recorded in the study that carries it, together with the number it would change.

\subsection{What we would do next}

The sweep re-grades archived findings under four rules and shows that the auditor contrast
depends on the rule (\S\ref{sec:auditor}). Re-grading only moves a cut through findings already
produced, so it cannot bring every family to a matched operating point: one family has no point
under the four deterministic rules between 3.3\% and 34.0\% false positives. The experiment this
paper most needs is to re-ask, instructing each family at graded severity and reading afresh,
which can place the families at a common false-positive rate and decide whether the auditor
contrast survives there. Matching the sampling parameters across families would remove the
confound above, and reading the cross-vendor family past eight times would turn a statement about
our budget into one about a curve.

We ran a first version of the specification experiment, holding the code and the visible tests
fixed and varying only the specification (\S\ref{sec:inreview}). Two gaps remain, and they decide
what it means. The clarifier inferred the missing rule from failing inputs and was wrong on most
instances, so the rule's correctness was measured, not manipulated. Writing the rule from the
oracle for some instances, and a deliberately wrong rule of matched length for others, would
separate a specification that settles the question from one that settles it correctly. Its
manipulation check also did not show the placebo inert, and a placebo rated no more determining
than the original would. Both are cheap, because the code, the tests and the adjudication rubric
already exist.

Beyond those: adjudicating more than one reading of the shipped auditor would let a naming rate be
compared with a union flag rate at the same $K$. A second generator would show whether the
residual's character belongs to the benchmark or to one model's failures. A population of natural
defects independently known to be specification-determined would separate the injected
population's artificiality from the property it was built to isolate. And a substrate whose ground
truth is not a hidden test suite would test the assumption every ceiling here rests on, that a
test suite asked the right question.

\section{Limitations in full}
\label{sec:limitations-full}
\textbf{Substrate.} The general-code studies measure function-level Python tasks from two code
benchmarks, and the science studies one scientific-code benchmark, small results artefacts built from it, and four tabular
datasets with synthetic faults. Nothing here establishes that the same ceiling, or the same
explanations, hold for another language, for repository-scale changes, for code whose correctness is
not decided by an executable suite, or for scientific artefacts larger than a model can read whole.
The results and data studies' faults are injected; the code study's failures are the generator's
own, judged by a benchmark that contains defective specifications and tests
\citep{hu2026scicodeverified}. One generator wrote every program. The code and data studies read two
auditor routes, the results study one, each at its default sampling.

\textbf{The ground truth is a test suite, and a test suite is not a specification.} Hidden
suites encode a reference implementation's behaviour, including behaviour its prose never fixed.
In the intervention study reported in \S\ref{sec:inreview}, a generator shown an instance's
failing inputs inferred a behavioural rule that \emph{contradicted} the hidden oracle on 23 of
32 instances. The study cannot say whether that reflects the specifications or the generator's
inference from inputs alone; either way it bounds what any recall figure measured this way can
mean.

\textbf{Interval coverage, measured only in part.} The primary intervals are percentile
bootstraps over problem clusters (over items within dataset in the data study). For a \emph{single rate} only, simulated under the design's
own cluster profiles and an assumed within-problem correlation, with 600 resamples rather than
the 10{,}000 the analyses use (400 replicates; Monte Carlo standard error 0.01--0.02), a nominal
$95\%$ interval covers between $0.93$ and $0.97$ at the tested true rates from $0.10$ to $0.60$, and $0.80$ at a rate of $0.03$ on the
56-cluster stratum (\S\ref{sec:stats}), which suggests that small-rate intervals may be narrower
than they claim. Two earlier interval constructions in this programme were found by review to cover $0.416$ and $0.075$ where $0.95$
was claimed.

\textbf{The validation instrument shares the failure mode under study.} Reports were checked by a
language model, which is what this paper measures a ceiling on. Studies cleared review at between
three and twenty-one rounds, with the round before the last always returning substantive
findings. Measurements moved on review in three of the eight general-code studies and in the post hoc
counts of two science studies; elsewhere the refusals concerned
interpretation, where the instrument is weakest. A further pass under an \emph{attempted}
conclusion-suppressed prompt produced six substantive findings that six briefed passes had missed,
but the supplied artefacts exposed the conclusions anyway, so blinding was not achieved and the
comparison is uncontrolled. The defect class it
leaves open is the one that looks right to a language model, and we do not know what remains in
it.

\textbf{The reviewer is one of the measured families.} \texttt{gpt-6-astra} reviewed every
report in this programme and is also an auditor route whose ranking against the others the paper
reports. For claims about \texttt{astra}'s relative performance the validation is a
self-assessment, and those comparisons should be treated as bracketed until a model that audited
nothing here has checked them.

\textbf{No human rater.} Every rating and adjudication in this work was made by a model, one of
them the agent that designed the studies and wrote the rubrics (\S\ref{sec:methods-review}). A
third model rating does not discharge that, and the design that asks for an outside human rater
has not been run.

\textbf{Budget.} Reading depth $K$, the number of arms, and the size of every intervention
population were set by cost rather than by a power calculation, and no sample size was justified
in advance by a minimum-detectable-effect argument.

\section{Related work in full}
\label{sec:related-full}

\textbf{Critics that read code.} Trained critic models write natural-language critiques of
model-written code and catch more bugs than unaided humans \citep{mcaleese2024critics}. That work
evaluates on both deliberately inserted and naturally occurring bugs. On code containing natural
model errors its critiques are preferred to human critiques in 63\% of cases, and it surfaces
errors in training data previously rated flawless. It also reports that critics hallucinate bugs,
and that a human--model team hallucinates less than the model alone. The natural-defect setting
and the precision cost of a louder critic are therefore already its territory, and we claim
neither. Its reported quantity is a preference or a per-critique comparison. What we add on this
axis is narrow: the union of repeated independent readings as an instrument for locating a limit,
ground truth supplied by an interpreter, and the false-positive rate carried beside every recall
figure. Aggregating several review passes is known to help. On 1{,}000 verified pull requests,
multi-review aggregation raises F1 by up to 43.67\% \citep{zeng2025swrbench}. The motivation is
shared with our union-of-$K$, but there an additional model call synthesises the sampled reviews,
while our union counts an instance whenever any reading flags it, so the two estimate different
things.

\textbf{Repeated sampling and coverage.} Coverage, the fraction of problems solved by at least one
of $K$ samples, rises steeply and then flattens. This is an established regularity of repeated
sampling, reported over four orders of magnitude of $K$ and fitted by a power law
\citep{brown2024monkeys}. Our union-of-$K$ is that quantity with a flag rule in place of a
correctness check, so \S\ref{sec:repeat} re-measures a known shape, and at $K \le 8$ our budget is
small beside theirs. What is ours is the reading: the curve is taken as a measurement of a
ceiling, reported beside the false-positive rate it costs, and tested against a registered
flattening bar that it fails, which is why we report diminishing returns and no limit.

\textbf{Self-assessment.} Two lines of work bear on a model judging its own output. Self-correction
without external feedback does not reliably improve reasoning and can degrade it
\citep{huang2024selfcorrect}. Models used as judges prefer their own generations, mostly
legitimately, since stronger generators are better judges, but harmfully on the instances they
themselves get wrong \citep{chen2025selfpreference}. Our same-vendor arms are the code-audit case of
that literature; because routes differ in model, operating point and sampling, we report their
contrasts as contrasts between routes, with the false-positive rate beside the recall.

\textbf{What the hidden suite decides.} EvalPlus rebuilt HumanEval and MBPP with larger suites and
reported defects in the original reference solutions \citep{liu2023evalplus}; its code carries
bespoke oracles for tasks where differential testing against the reference did not apply.
\citet{richter2026collapse} show that models collapse onto a single incorrect reading of an
underspecified task, and name twelve MBPP tasks they identified by hand. Eleven fall inside our
defect population, and our raters, who labelled them before that paper was fetched, agree on nine.
On the test-generation side, the dominant error in generated tests is a wrong expected value
rather than an invalid input \citep{xiong2023testing}, and generated oracles tend to capture what
a program does rather than what was intended \citep{konstantinou2024oracles}. Our claim is the
auditing counterpart: where the prose does not fix the value, a recall number computed against
that suite measures the benchmark as much as the auditor.

\textbf{Consensus as a substitute for ground truth.} Several systems validate generated tests
without a reference by agreement, through self-consistency, dual execution, or multi-agent
discussion for {JUnit} suites \citep{taherkhani2026consistency, xu2025consensus}. Both measure what
survives the filter against a ground-truth implementation: 84--91\% of retained tests valid in one
\citep{taherkhani2026consistency}, 0.89--0.93 oracle correctness in the other
\citep{xu2025consensus}. Our measurement differs in unit and in bar. It is the share of wrong tests
among retained \emph{failing test--candidate applications}, against a preregistered 2\%
threshold, over a frozen set of generated tests and candidates that includes candidates already
failing their visible tests. Corroboration across independent generations does not reach it
(\S\ref{sec:inreview}), and wrong tests recur by problem. That recurrence is compatible with the
self-consistent errors, the same wrong answer produced again, that degrade the detection methods
\citet{tan2025consistent} evaluate. Our records hold indices, not assertions, so they
do not show that the recurring errors are the same one.

\textbf{Auditing scientific work.} On manuscripts, no model exceeds 21.1\% recall or 6.1\% precision
on errata-grade errors and repeated runs rarely rediscover the same errors \citep{son2025spot}; the
best models detect 46.7\% of real paper--code discrepancies \citep{baumgartner2026scicoqa}. An LLM
auditor of AI-scientist projects detects hand-labelled pitfalls far better with logs and code than
with the paper alone \citep{luo2025pitfalls}; pattern-based linting targets methodology bugs in
scientific Python \citep{samsonau2026scicodelint}; cleaning agents find row-level errors in corrupted
tables and miss errors spanning rows \citep{bendinelli2025cleaning}; and simulation code can run and
still solve the wrong equations \citep{song2026moose}. SciCode supplies scientist-written tests
\citep{tian2024scicode}, and an expert audit of it found many specifications and tests that reject
correct solutions \citep{hu2026scicodeverified}; reproduction benchmarks ask agents to recover
reported results \citep{siegel2024corebench}. On constructed statistical workflows, LLM reviewers
detect planted flaws almost always but flag 17--50\% of correct scripts \citep{chelate2026replicationtrap},
and Luo et al.\ already report both true- and false-positive rates. Scientific auditing and paired
operating points are therefore not new. What we add is this measurement programme applied to
scientific code, and the comparison of the LLM tier with deterministic verification on reported
results and data, which we did not find in prior work.

\textbf{Who audits whom.} That an evaluator should not be the model under evaluation is supported
by measurements of self-preference in LLM judges against verifiable ground truth
\citep{chen2025selfpreference}. We do not measure self-preference. A frozen generator's code
audited by routes that differ in model, operating point and sampling cannot separate preference
from performance, so we report the same-vendor contrast as a contrast between routes, on one
substrate.

\textbf{Preregistration.} Preregistering an evaluation protocol is beginning to appear in machine
learning, both as a protocol for experiments with agents \citep{vaccaro2026prereg} and as a defence
against tuning prompts, decoding parameters or output format until a desired result appears
\citep{thomas2026phacking}. We count our use of it as method, not contribution.

\section{Reproducibility and claims}
\label{sec:repro}
\ifdefined\ANON An anonymised repository (link withheld for review) carries the{} \else
The project repository, to be released with a later version of this paper, carries the{} \fi
claims ledger (\texttt{manuscript/CLAIMS.md}), with the claims the evidence does not support; the results
report of every study the paper draws on; every registered study's preregistration and numbered
amendments (the science studies' under \texttt{plan/studies/ai4s/}, where the data and results
registrations carry errata); every independent review of those reports, verbatim and including
refusals; and the derived records (counts, rater labels, hashes and intervals) that the figures
and Table~\ref{tab:primary} are drawn from. A single command, \texttt{reproduce.sh},
regenerates Figures 1--4 and Table~\ref{tab:primary} from those records and checks that each
is byte-identical to the copy the manuscript compiles, with no model calls, no network access and no path outside the checkout; it was run from
a fresh clone with the home directory unavailable to establish that. The
repository's \texttt{README} maps each claim to its report, its reviews and its records. The
corpora are HumanEval (MIT), HumanEval+ and MBPP+ (Apache-2.0), SciCode (Apache-2.0) and four UCI
datasets (CC BY 4.0), which permit redistribution. What a replication cannot do from this repository alone is regenerate the
records themselves. With the archive of raw readings, which is kept separately, they regenerate
deterministically through the analysis code in the auditor harness, a separate repository named
in the provenance file; without it, only by re-running the studies against model endpoints whose
outputs are not deterministic.

\end{document}